\documentclass[amsmath,12pt,amssymb,preprint,prd,aps,nofootinbib]{revtex4}
\usepackage{amsfonts} 
\usepackage{graphicx} 
\usepackage{epsfig}
\usepackage{multirow}
\usepackage{xcolor}
\usepackage{bm}

\newcommand{\diracslash}[1]{#1\llap{/\kern2pt}}

\newcommand{\be}{\begin{equation}}
\newcommand{\ee}{\end{equation}}
\newcommand{\bea}{\begin{eqnarray}}
\newcommand{\eea}{\end{eqnarray}}
\newcommand{\ba}[1]{\begin{array}{#1}}
\newcommand{\ea}{\end{array}}

\newcommand{\bt}{\begin{tabular}}
\newcommand{\et}{\end{tabular}}

\newcommand{\beas}{\begin{eqnarray*}}
\newcommand{\eeas}{\end{eqnarray*}}

\begin{document}
\title{Thermodynamic and transport properties of 
hot asymmetric nuclear matter within a chiral SU(3) model}
\author{Amruta Mishra}
\email{amruta@physics.iitd.ac.in}
\affiliation{Department of Physics, Indian Institute of Technology Delhi, 
Hauz Khas, New Delhi -- 110016, India}
\author{J. Schaffner-Bielich}
\email{schaffner@astro.uni-frankfurt.de}
\affiliation{Institut f\"ur Theoretische Physik,
Goethe Universit\"at \\ Max von Laue Strasse 1,
D--60438, Frankfurt am Main, Germany}
\vskip -1.5in
\begin{abstract}
We investigate the thermodynamic and transport properties in hot nuclear
matter accounting for the medium modifications of the nucleons within a chiral
SU(3) model including effects from isospin asymmetry. Using the relaxation
time approximation, the transport coefficients of the shear viscosity and
thermal conductivity are studied. The shear viscosity, $\eta$, calculated within
the chiral SU(3) model is observed to be smaller than the values calculated
for free nucleon gas, whereas the thermal conductivity $\kappa$ 
is appreciably larger as compared to the free nucleon gas. 
The presence of isospin asymmetry in the medium leads to higher 
values of both the coefficients of shear viscosity ($\eta$)
and thermal conductivity ($\kappa$), however, the effect is observed to be
marginal for $\eta$. In the chiral SU(3) model, the effect of 
isospin asymmetry is observed to be larger for higher values 
of temperature. For T=150 MeV, there is observed to be a drop in
the value of $\kappa$ as density is increased, contrary to the 
increase observed for the lower values of temperature, T=50 and 100 MeV.
The shear viscosity coefficient to entropy density ratio 
$\eta/s$ drops with increasing baryon density that becomes
more pronounced at higher temperatures in the chiral SU(3) model as compared
to the case of a free nucleon gas. 
The present study of the thermodynamic as well
as transport properties in hot nuclear matter is of relevance for relativistic
heavy-ion collisions with different initial isospin asymmetry, in particular
for the compressed baryonic matter experiment at the FAIR facility at
GSI.
\end{abstract}
\maketitle
\newpage

\section{Introduction}
\label{intro}
The topic of study of the effects of the temperature
and/or density on the properties of QCD matter
is an interesting and important area
of research due to its relevance in relativistic 
heavy-ion collision experiments. The 
hot and dense strongly interacting matter created in the
high energy nuclear collision experiments can modify 
the experimental observables, e.g. the particle yields,
particle spectra and collective flow of the particles. 
The transport properties, e.g., the shear and
bulk viscosity coefficients, are important to study
as they can affect the collective flow coefficients, and
also, the temperature and chemical potential
dependence of these coefficients can give information 
regarding the phase transitions in the QCD phase diagram.
These transport properties have been studied extensively 
in the literature within the framework of relativistic kinetic theory,
both in hadronic \cite{Danielewicz_PLB_146_168_1984,Prakash_Phys_Rep_227_321_1993,Itakura_PRD77_014014_2008,Recent_Prog_QHD_Serot_Walecka_IJMPE6_515_1996,NPA573_554_1994_Monras_Trans_coeffs_Nucl_Neutron_matter,Hakim_Mornas_PRC47_2846_1993,Ayik_Ivanov_Russkikh_Norenberg_NPA578_640_1994,Abu_Samreh_NPA552_1993_101,A_S_Khvorostukhin_NPA915_198_2013,Chakraborty_Kapusta_PRC83_014906_2011,Albright_Kapusta_PRC93_014903_2016,PRC86_024913_2012,PRC77_024911_2008_Gorenstein}
as well as in quark-gluon plasma (QGP) phases 
\cite{Phys_Rev_D31_53_1985_Danielewicz_Gyulassy,Hosoya_Kajantie_NPB250_666_1985,PRC84_035202_2011_visc_gluon_matter,Gavin_NPA435_1985_826,A_S_Khvorostukhin_NPA845_106_2010,Phys_Rev_C103_054901_2021_Elena,2408_00524_Isabella_Moore}.
At low collision energies, when the system consists of hadrons,
the dynamics of the heavy-ion collisions
is described well using the hadronic transport approaches.
However, at high collision energies, 
the hybrid approaches are used that incorporate
the relativistic viscous hydrodynamics for the QGP phase 
along with the hadronic transport theory 
to describe the dynamics of the heavy-ion collisions
\cite{Hybrid_HIC}. The effect of the baryon densities 
and temperatures on the shear viscosity to entropy density ratio, $\eta/s$ 
have been studied using the hybrid approaches
\cite{Hybrid_HIC_shear_1,Gotz_Hanah_PRC106_054904_2022,Gotz_Hanah_2503_10181}.
The AdS/CFT analysis of the shear viscosity leads
to a lower bound of $\eta/s$ as $1/4\pi$ \cite{KSS}. 
The particle flow anisotropies at the Relativistic Heavy Ion
Collider (RHIC) are explained well 
in ideal hydrodynamics 
\cite{particle_flow_anisotropies_1,particle_flow_anisotropies_2,particle_flow_anisotropies_3}
from which it is inferred that the produced matter at RHIC
is close to an ideal fluid, with small value of $\eta/s$. 
There have been studies for the transport coefficients
using the hadron resonance model including excluded volume 
\cite{PRC86_024913_2012,PRC77_024911_2008_Gorenstein}, 
as well as using the mean field Walecka model with the nucleons
interacting via scalar and vector interactions
\cite{Itakura_PRD77_014014_2008,Recent_Prog_QHD_Serot_Walecka_IJMPE6_515_1996,NPA573_554_1994_Monras_Trans_coeffs_Nucl_Neutron_matter,Hakim_Mornas_PRC47_2846_1993,Ayik_Ivanov_Russkikh_Norenberg_NPA578_640_1994,Abu_Samreh_NPA552_1993_101,A_S_Khvorostukhin_NPA915_198_2013,Chakraborty_Kapusta_PRC83_014906_2011,Albright_Kapusta_PRC93_014903_2016}.
The transport coefficients in hot nuclear matter
have been studied by solving the Boltzmann equation for the distribution
function of the particle (antiparticle), whose derivative with respect
to time (which is zero for equilibrated matter) arises from the collision 
integral. The viscosity and thermal conductivity coefficients have been studied 
in first order Chapman-Enskog expansion using the Boltzmann-Uehling-Uhlenbeck 
collision term, incorporating the Pauli blocking factors
in Refs. \cite{Danielewicz_PLB_146_168_1984,Abu_Samreh_NPA552_1993_101}.
The transport coefficients have also been studied extensively 
in the literature accounting for the in-medium
effects for the nucleons via interactions with scalar and vector mesons
in the Walecka model with the Boltzmann-Uehling-Uhlenbeck (BUU) 
collision term \cite{BUU_1,BUU_2,BUU_3,BUU_4,BUU_5,BUU_6,BUU_7,BUU_8} 
as well as in the relaxation time
approximation. The results for the transport coefficients
of hot nuclear matter calculated using the Walecka model
with BUU collision term 
\cite{NPA573_554_1994_Monras_Trans_coeffs_Nucl_Neutron_matter},
are observed to be in good agreement with the results arising from
the relaxation time approximation 
of Ref. \cite{Hakim_Mornas_PRC47_2846_1993}. The latter  
is mathematically a much simpler problem and the above agreement
shows that this approximation can describe well the transport properties
of hot hadronic matter. 
The transport coefficients (shear and bulk viscosities and
thermal conductivity) have been derived in a quasiparticle
approximation for the nucleons interacting with the scalar and vector 
mesons in a quantum hadrodynamic framework in Ref. 
\cite{Albright_Kapusta_PRC93_014903_2016}. 
In the present work, we study the 
transport coefficients of hot asymmetric 
nuclear matter in the relaxation time approximation,
incorporating the in-medium effects for the nucleons 
calculated within a chiral SU(3) model.
The effects of
the isovector-scalar ($\delta$) and isovector-vector ($\rho$) mesons
are taken into account for the study of nucleons 
in asymmetric nuclear matter, in addition to the isoscalar-scalar 
and isoscalar-vector mesons. The chiral SU(3) model 
adopted in the present work 
\cite{paper3,hartree,kristof1}
describes well the properties of the nuclear matter, finite nuclei
and neutron stars \cite{Schramm_2013,Dex_2015,JSB_2016}.
The model has also been generalized to study heavy
flavor (charm and bottom) mesons in (magnetized) hadronic matter
\cite{AMSPMWG_2015,AMSPM_2017,AMSPM_DW_HQ_DS_PV_2023,AMAKSPM24}. 
Within the model, the baryon masses are generated from 
spontaneous chiral symmetry breaking, with the scalar mesons
(proportional to the light quark condensates) attaining 
vacuum expectation values. The nucleon masses
in the hot asymmetric nuclear medium 
are obtained from the mean values of the
scalar fields ($\sigma\sim \langle \bar u u + \bar d d\rangle$,
$\zeta\sim \langle \bar s s \rangle$ and
$\delta\sim \langle \bar u u - \bar d d\rangle$),
which are solved from their equations of motion.
To study the effects of (partial) chiral symmetry restoration,
which leads to the mass drop of the nucleons in the nuclear medium,
on the transport properties is the motivation of the present work.

The transport properties of pion gas and nuclear matter,
as well as hadronic mixtures (pion-nucleon, pion-kaon-nucleon) 
have been studied in the literature \cite{Prakash_Phys_Rep_227_321_1993,
Itakura_PRD77_014014_2008}. 
In Ref. \cite{Itakura_PRD77_014014_2008},
a study of a mixture of pions and nucleons shows that 
for large baryon chemical potential, $\mu$ and moderate temperature (T),
the shear viscosity coefficient is dominated by nucleon contributions,
whereas, small $\mu$ and large T is dominated by pion contributions.
The present study of transport properties of hot hadronic
matter using a chiral SU(3) model can be of relevance
for the Compressed Baryonic matter (CBM) experiment 
at FAIR at the future facility at GSI
\cite{TheCBMPhysicsBook,P_Senger_HIC_FAIR_NICA_Energies},
as well as beam energy scan at RHIC \cite{BES_RHIC}, 
which aim to produce matter at high densities and moderate 
temperatures; 
hence the transport properties are expected
to be dominated by the nucleons.
These transport properties can have implications on the collective flow
coefficients as well as hadron spectra. 
The isospin asymmetry effects on the thermodynamic
and transport properties of the hot nuclear matter, 
which are observed to be appreciable at high densities, 
can be relevant for asymmetric heavy-ion collision experiments 
planned at CBM at the future facility at GSI.
The difference in the spatial density distributions of the protons
and neutrons of colliding nuclei (due to $N\ne Z$)
leads to isospin asymmetry in the system. To study the 
effects of isospin asymmetry on the thermodynamic and 
transport properties of the hot hadronic medium
in the present work, we choose the value of the isospin 
asymmetry parameter $\eta_N\equiv(\rho_n-\rho_p)/\rho_B$ to be $0.6$.
For this value of $\eta_N$, the effect of isospin asymmetry 
is observed to lead to quite significant increase in the
coefficient of thermal conductivity $\kappa$, especially
at high temperature, whereas the effect on the coefficient 
shear viscosity $\eta$ is observed to be quite small.
The present study of the effects of the isospin asymmetry,
however, is not relevant for the neutron stars, 
which comprise of highly isospin asymmetric 
matter, as the effects of the shear viscosity and thermal conductivity
do not play an important role unless neutrinos are trapped 
\cite{Alford_NS}.

The paper is organized as folllows. In Sec. II, 
the chiral SU(3) model used for computing the transport 
coefficients of hot hadronic matter is described briefly. 
The thermodynamic properties of hot asymmetric nuclear
matter are studied using the model.
Sec. III decribes the derivation of the transport coefficients
by solving the Boltzmann equation in the relaxation time 
approximation, accounting for the medium modifications 
of the nucleons as obtained using the chiral SU(3) model. 
In Sec. IV, we describe
the results obtained for the thermodynamic quantities 
and transport coefficients, namely the shear viscosity
coefficient, and the thermal conductivity.
Sec. V summarizes the findings of the present work.

\section{Chiral SU(3) model}

In this section, we briefly describe the chiral SU(3) model 
\cite{paper3,hartree,kristof1,AMAKSPM24} used to study
the thermodynamic properties and the transport properties
of hot nuclear matter in the present study.
The model \cite{paper3,hartree,kristof1}
is based on a nonlinear realization of chiral symmetry 
\cite{Weinberg,coleman,Bardeen} 
and incorporates the broken scale invariance
of QCD through the introduction of a scalar dilaton field, $\chi$
\cite{sche1,heide1}. 
The general form of the Lagrangian density of chiral SU(3) model 
is written as \cite{paper3}
\begin{equation}
{\cal L} = {\cal L}_{kin}+\sum_{W=X,Y,V,A,u} {\cal L}_{BW} + 
{\cal L}_{vec} + {\cal L}_{0} + {\cal L}_{scale-break}+
{\cal L}_{SB}.
\label{genlag}
\end{equation}
In Eq.(\ref{genlag}), ${\cal L}_{kin}$ is the kinetic energy term, 
${\cal L}_{BW}$ is the baryon-meson interaction term, where, 
$W=X,Y,V,A,u$ correspond to the interactions of the baryons with
the scalar, pseudoscalar singlet, vector meson, axial-vector meson 
and pseudoscalar octet mesons respectively.
The baryon-spin-0 meson interaction term generates the baryon masses.
${\cal L}_{vec}$  describes 
the dynamical mass generation of the vector mesons via couplings to the 
scalar mesons and contain additionally quartic self-interactions of the 
vector fields. ${\cal L}_{0}$ contains the meson-meson interaction terms, 
${\cal L}_{scale-break}$ is the QCD scale breaking term 
in the Lagrangian density, 
including a logarithmic potential of the scalar dilaton field, $\chi$,
and, ${\cal L}_{SB}$ describes the explicit chiral symmetry breaking.
In the mean field approximation, the meson fields are treated as
classical fields, and, the meson fields that have nonzero contributions
to the baryon-meson interactions are the scalar and the vector mesons,
with the interaction Lagrangian given as
\begin{equation}
{\cal L}_{BS}+ {\cal L}_{BV}=\sum _i {\bar  \psi}^i
(g_{\sigma i} \sigma+g_{\zeta i} \zeta+g_{\delta i} \delta
-g_{\omega i}\gamma_\mu\omega^\mu
-g_{\rho i}\gamma_\mu {\vec \tau} \cdot {\vec {\rho^{\mu}}})\psi^i,
\label{BS_BV}
\end{equation}
where, $i$=p,n for nuclear matter as considered in the present work,
$\sigma$ and $\zeta$ are the scalar-isoscalar nonstrange and strange
fields, $\delta$ is the isovector-scalar fiel, and
$\omega^\mu$ and $\vec {\rho^{\mu}}$ are the isoscalar-vector
and isovector-vector fields.
For uniform and rotationally invariant matter, the expectation
values of the meson fields are independent of space-time. Also,
the spatial components of the vector fields have zero expectation 
values, i.e., for the vector fields, 
$\omega^\mu \rightarrow \delta^{\mu 0}\langle \omega^0\rangle 
\equiv \delta^{\mu 0}\omega$ and 
$\rho^{\mu a} \rightarrow \delta^{\mu 0} \delta^{a 3}
\langle \rho^{0 3}\rangle 
\equiv \delta^{\mu 0} \delta^{a 3} \rho$.
The effective mass of the nucleon of species 
$i$ is obtained from the in-medium values of the 
scalar mesons and is given as
\begin{equation}
m_{i}^{*} = - g_{\sigma i}\sigma- g_{\zeta i}\zeta - g_{\delta i} \delta,
\label{ameff}
\end{equation}
and the effective chemical potential for the $i$-th nucleon, 
due to the interaction with the vector mesons ($\omega$ and $\rho$)
is given as
\begin{equation}
\mu^*_i=\mu_i-g_{\omega i}\omega-g_{\rho i}\rho.
\label{amustr}
\end{equation}
The other terms in the Lagrangian density of equation (\ref{genlag}) 
are given as
\begin{align}
{\cal L}_{vec} &= \frac{1}{2} \frac{\chi^2}{\chi_0^2}\Big(
m_{\omega}^{2} \omega^ 2+m_{\rho}^{2} \rho^ 2
\Big) +g_4 (\omega^4 
+6 \omega^2 \rho^2+\rho^4)\\
\label{L_vec}
{\cal L}_{0} & = - \frac{ 1 }{ 2 } k_0 \chi^2
(\sigma^2+\zeta^2+\delta^2) + k_1 (\sigma^2+\zeta^2+\delta^2)^2
\nonumber \\
     &+ k_2 ( \frac{ \sigma^4}{ 2 } + \frac{\delta^4}{2} + \zeta^4
 +3 \sigma^2 \delta^2)
     + k_3 \chi (\sigma^2 - \delta^2) \zeta - k_4 \chi^4 \\
\label{L_0}
{\cal L}_{scale \; break}&= -\frac{1}{4} \chi^{4} 
{\rm ln} \frac{\chi^{4}}{\chi_{0}^{4}} + \frac{d}{3} \chi^{4} 
{\rm ln} \Bigg( \frac{\left( \sigma^{2} - \delta^{2}\right)\zeta }
{\sigma_{0}^{2} \zeta_{0}} \Big( \frac{\chi}{\chi_{0}}\Big) ^{3}\Bigg), \\
\label{scalebreak}
{\cal L} _{SB} & =  - \left( \frac{\chi}{\chi_{0}}\right) ^{2} 
\left[ m_{\pi}^{2} 
f_{\pi} \sigma + \left( \sqrt{2} m_{K}^{2}f_{K} - \frac{1}{\sqrt{2}} 
m_{\pi}^{2} f_{\pi} \right) \zeta \right]. 
\end{align}
In the above, the parameters $k_0$, $k_2$ and $k_4$ are fitted
to ensure extremum in the vacuum for the equations of  motion
for scalar fields $\sigma$, $\zeta$ and the dilaton field $\chi$, 
$k_1$ is fitted to reproduce
the mass of $\sigma$ to be of the order of 500 MeV, $k_3$
is fitted from the $\eta$ and $\eta'$ masses, and the value of
the $\chi$ in vacuum is fitted so that the pressure $p=0$ 
at the nuclear matter saturation density.
The thermodynamic potential per unit volume for the hadronic medium 
is given as
\begin{eqnarray}
 && \Omega/V  = - {\cal L}_{vec} - {\cal L}_0 - {\cal L}_{scale\;break}
- {\cal L}_{SB} -{\cal V}_{vac} \nonumber \\
&& -T\sum_i \gamma_i \int \frac{d^3 {\bf p}}{(2\pi)^3} 
 \Big[
\ln \Big( 1+e^{-(E_i^*({\bf p})-\mu_i^*)/T} \Big)+
\ln \Big( 1+e^{-(E_i^*({\bf p})+\mu_i^*)/T} \Big) \Big],
\label{thermo_total}
\end{eqnarray}
where, $\gamma_i$=2 is the spin degeneracy factor for the 
$i$-th baryon ($i$=p,n for nuclear matter as considered 
in the present work), 
$E_i^*({\bf p}) =({{\bf p}^2+m_i^*}^2)^{1/2}$
is the single particle energy of nucleon of species $i$, 
with the effective mass and effective chemical
potential given by equations (\ref{ameff})
and (\ref{amustr}) respectively. 
The potential at $\rho_B$=0 and $T=0$, 
${\cal V}_{vac}(=-{\cal L}_{vac})$ has been 
subtracted to get a vanishing vacuum energy.
The thermodynamic potential per unit volume 
of the hot nuclear matter is related
to the thermodynamic quantities pressure, energy density and
entropy density as 
\begin{equation}
p=-\Omega/V=-\Big(\epsilon-Ts-\sum_i \mu_i \rho_i\Big),
\label{pressure}
\end{equation}
with the expressions for the entropy density, $s$
and energy density, $\epsilon$ given as
\begin{eqnarray}
s & = & 
-\sum_i \gamma_i \int \frac{d^3 {\bf p}}{(2\pi)^3} 
\Big [f_i^{eq}({\bf p}) \ln f_i^{eq} ({\bf p})
+(1-f_i^{eq}({\bf p})) \ln (1- f_i^{eq} ({\bf p}))\nonumber \\
&+& {\bar f_i}^{eq}({\bf p}) \ln {\bar  f_i}^{eq} ({\bf p})
+(1-{\bar f_i}^{eq}({\bf p})) \ln (1- {\bar  f_i}^{eq} ({\bf p}))
\Big]
\label{entr}
\end{eqnarray}
and,
\begin{eqnarray}
\epsilon 
&  = &  
\sum_i \gamma_i \int \frac{d^3 {\bf p}}{(2\pi)^3} 
E_i^*({\bf p})
\Big [f_i^{eq}({\bf p})+{\bar f}_i^{eq}({\bf p})\Big]
- {\cal L}_{vec} - {\cal L}_0 
\nonumber \\
& - &  {\cal L}_{scale\;break}
 - {\cal L}_{SB} -{\cal V}_{vac} 
+(g_{\omega i}\omega+g_{\rho i}\rho) \rho_i, 
\label{energy_density}
\end{eqnarray}
where, $\rho_i$ is the number density 
of the $i$-th baryon, given by the expression
\begin{align}
\rho_{i}=  
\label{rho_i}
\gamma_i \int \frac{ d^3 {\bf p}}{(2\pi)^3} 
\Big [f_i^{eq}({\bf p})-{\bar f}_i^{eq}({\bf p})
\Big].
\end{align}
In the above equations, $f_i^{eq} ({\bf p})$ and ${\bar f_i}^{eq} ({\bf p})$
are the particle and antiparticle distribution functions 
in thermal equilibrium for nucleon of species $i$=p,n, given as
\begin{equation}
f_i^{eq}({\bf p})= \frac {1}{e^{{(E_{i}^*({\bf p})-\mu^{*}_i)}/{T}} +1},\;\;
{\bar f_i}^{eq}({\bf p})
= \frac {1}{e^{{(E_{i}^*({\bf p})+\mu^{*}_i)}/{T}} +1}.
\label{equil_distr_fns}
\end{equation}
For given values of the baryon density, $\rho_B=\sum_{i={\rm p,n}} \rho_i$,
the isospin asymmetry parameter $\eta_N$, and 
temperature $T$ of the hot nuclear matter, the values of the scalar fields 
($\sigma,\zeta,\delta$), the dilaton field $\chi$, 
and the vector fields ($\omega$ and $\rho$) are solved from
their coupled equations of motion obtained through minimization
of the thermodynamic potential. 

\section{Transport Coefficients}

In the present work, we study the coefficients of
shear viscosity and thermal conductivity 
of hot asymmetric nuclear matter using the chiral SU(3) model
described in the previous section. 
In Ref. \cite{Albright_Kapusta_PRC93_014903_2016}, 
the expressions for the transport coefficients 
of hot and dense hadronic matter 
have been derived using the quantum hadrodynamical model
describing the interactions of the nucleons with
a scalar ($\sigma$) and a vector ($\omega$) meson. 
We generalize the approach 
used in Ref. \cite{Albright_Kapusta_PRC93_014903_2016} to 
include the interactions of the nucleons with the scalar mesons
($\sigma$, $\zeta$ and $\delta$) and the vector mesons
($\omega$ and $\rho$) described by the Lagrangian
density given by equation (\ref{BS_BV}) 
within the chiral SU(3) model. 
The model includes the self-interactions of the meson fields 
given by equations (\ref{L_vec}) and (\ref{L_0}), 
additionally including the interactions of the dilaton field 
(given by equation (\ref{scalebreak})), which simulate 
the broken scale invariance of QCD. 
For uniform hot nuclear matter in thermal and kinetic equilibrium, 
described in the previous section using the chiral SU(3) model, 
the scalar and vector fields have space-time-independent 
expectation values. Also, the expectation values of the spatial components 
of the vector fields vanish due to rotational 
invariance in the rest frame of the nuclear medium
($u^\mu=(1,{\bf 0})$). However, for out-of-equilibrium nonuniform
matter with flow velocity $u^\mu=(1, {\bf v})$, the mean fields 
are no longer space-time independent, and, the spatial
components of the vector fields are no longer zero.
The dispersion relation for the $i$-th nucleon (antinucleon)
is then given in terms of the kinetic momentum
${\bf p}^*_i={\bf p} \mp (g_{\omega i}
{ \mbox{\boldmath $\omega$}} 
+g_{\rho i} {\mbox{\boldmath $\rho$}})$ as 
\begin{equation}
E_i ({\bf p}_i)=E_i^*({\bf p}^*_i)
\pm (g_{\omega i}{\bf \omega} +g_{\rho i}{\bf \rho}),
\end{equation}
where $E_i^*({\bf p}^*_i)=({{\bf p}^*_i}^2+{m_i^*}^2)^{1/2}$
is the single particle energy.

The energy-momentum tensor and the baryon current are given as
\begin{equation}
T^{\mu \nu}=-pg^{\mu \nu}+
{\cal W} u^\mu u^\nu 
+\Delta T^{\mu \nu}
\label{energy-mom}
\end{equation}
and
\begin{equation}
J_B^{\mu}=\rho_B u^\mu +\Delta J_B^\mu,
\label{baryon_current}
\end{equation}
with the dissipative parts given as 
\begin{equation}
\Delta T^{\mu \nu}=\eta \Big(D^\mu u^\nu+D^\nu u^\mu
+\frac{2}{3}\Delta^{\mu \nu}\partial_\alpha u^\alpha\Big)
-\xi \Delta^{\mu \nu}\partial_\alpha u^\alpha
\label{energy-mom_diss}
\end{equation}
and
\begin{equation}
\Delta J_B^{\mu}=\kappa \Big (\frac{\rho_B T}{\cal W}\Big)
D^\mu\Big(\frac{\mu_B}{T}\Big),
\label{baryon_current_diss}
\end{equation}
where $D^\mu=\partial ^\mu -u^\mu D$, 
$D=u^\alpha \partial _\alpha$ and
$\Delta ^{\mu \nu}=u^\mu u^\nu -g^{\mu \nu}$.
In the above equations, $\epsilon$, $p$, $\rho_B$,
$\mu_B$, T and ${\cal W}(=\epsilon+p)$ 
are the energy density, pressure, baryon density, 
baryon chemical potential, temperature and enthalpy density
respectively, $\eta$ and $\xi$
are the coefficients of shear and bulk viscosity, 
and $\kappa$ is the coefficient of thermal conductivity.
The transport coefficients of viscosity and thermal conductivity 
are computed from the dissipative parts of the energy-momentum tensor 
and the baryon current, by using the departures of the distribution
functions from the local equilibrium distribution functions.

The Boltzmann equation for the distribution functions 
of species $i=p,n$ 
for nuclear matter as considered 
in the present work is given as
\begin{equation}
\frac{ d {F_i({\bf x}, {\bf p}^*_i, t)}}{dt}=C_i,
\label{Boltzmann_eqn}
\end{equation}
where, $F_i=f_i (\bar f_i)$ is the distribution function
of the particle (antiparticle).
Considering first order departures from local equilibrium,
these distribution functions are written in terms of 
function $\phi_i (x,p)$ as 
\begin{equation}
{F_i}=F_i ^{eq} (1+\phi_i (x,p)),
\end{equation}
where $F_i^{eq}$=${f_i} ^{eq}$ (${\bar {f_i}} ^{eq}$) 
is the distribution function of the $i$-th nucleon (antinucleon)
in local equilibrium. 
The function $\phi_i(x,p)$ is expressed in terms of the first order 
space-time derivatives of the flow velocity, $u^\mu$,
in the same tensorial form as the dissipative parts of 
the energy-momentum tensor and baryon current, given by
equations (\ref{energy-mom_diss}) and (\ref{baryon_current_diss});
i.e., $\phi_i(x,p)$ is written as
\begin{equation}
\phi_i(x,p)={\tilde A}_i \partial_\alpha u^\alpha
-{\tilde B}_i p^\mu_i D_\mu\Big(\frac{\mu_B}{T}\Big)
+{\tilde C}_i p^\mu_i p^\nu_i \Big(D_\mu u_\nu+D_\nu u_\mu
+\frac{2}{3}\Delta_{\mu \nu}\partial_\alpha u^\alpha\Big),
\label{phi_i}
\end{equation}
where ${\tilde A}_i$, ${\tilde B}_i$, ${\tilde C}_i$
are functions of $p$.
Including 2-2, 2-1 and 1-2 reactions and 
ensuring that the particles in these reactions are 
in chemical equilibrium in these reactions, 
the collision integral $C_i$
on the right-hand side of the Boltzmann equation
is computed 
retaining terms up to linear 
in $\phi_i (x,p)$. The functions ${\tilde A}_i$, ${\tilde B}_i$
and ${\tilde C}_i$ [and, hence, $\phi_i(x,p)$] are obtained
by solving the Boltzmann equation using the Landau-Lifshitz 
condition. From this, one obtains the departure of the distribution
function from the local equilibrium function,
$\tilde {\delta F_i}=F_i-F_i^{eq}$, as expressed in terms
of the nonequilibrium energy $E_i$, which is the quantity
that is conserved in local collisions.
Writing the dissipative parts of the energy-momentum tensor
and the baryon current in terms of $\tilde {\delta F_i}$,
the expressions of the coefficients of shear viscosity and 
thermal conductivity are derived.
In the classical approximation, the distribution function 
in local equilibrium $F_i^{eq}$ has the form 
\begin{equation}
F_i^{eq}\equiv F^{eq}_{i\,({\rm cl})}
=f^{eq}_{i\,({\rm cl})} ({\bar f}^{eq}_{i\,({\rm cl})})
=e^{-{{(u_\alpha (x) {p_i^*}^\alpha}\pm \mu^{*}_i)}/T}
\label{classical_distr}
\end{equation}
for the nucleon (antinucleon), which reduces to the expression
$F_i^{eq}=e^{-{(E_{i}^*({\bf p})\pm \mu^{*}_i)}/T}$
in the rest frame ($u^\mu=(1,{\bf 0})$) of the nuclear matter. 
The quantum statistical distribution function in local equilibrium
has the form given by equation (\ref{equil_distr_fns}) in the
rest frame of uniform nuclear matter,
which can be rewritten as 
\begin{equation} 
F_i^{eq}\equiv F^{eq}_{i\,({\rm quantum})}
=\frac{F^{eq}_{i\,({\rm cl})}}{1+F^{eq}_{i\,({\rm cl})}},
\label{quantum_distr}
\end{equation}
with $F^{eq}_{i\,({\rm cl})}=
f^{eq}_{i\,({\rm cl})} \big({\bar f}^{eq}_{i\,({\rm cl})}\big)$
corresponding to the nucleon (antinucleon).
However, it might be noted here that, unlike the case
of the distributions functions of equation 
(\ref{equil_distr_fns}) corresponding to the 
uniform nuclear matter at rest,
where, the effective energy and effective chemical potential
are independent of $x$, in the expression of
$F^{eq}_i$ in local equilibrium, these are dependent 
on $x$ due to the space-time-dependent scalar and 
vector mean fields. The coefficients of viscosity 
and the thermal conductivity
have been derived in Ref. \cite{Albright_Kapusta_PRC93_014903_2016},
in the classical approximation using the distribution
function given by equation (\ref{classical_distr}), 
as well as, using the approximate form of the quantum distribution 
function obtained by assuming $F^{eq}_{i\,({\rm cl})}$ to be small as 
compared to unity in equation (\ref{quantum_distr}), 
given as
\begin{equation}
F^{eq}_{i\,{\rm (quantum)}} \approx F^{eq}_{i\,{\rm(approx)}}=
F^{eq}_{i\,{\rm (cl)}}\big(1-F^{eq}_{i\,{\rm (cl)}}\big).
\label{quantum_distr_approx}
\end{equation}
In the energy-dependent relaxation time approximation, 
it is assumed that the particle (antiparticle) of nucleon 
species $i$ is out of equilibrium ($\phi_i (x)$ is nonzero) 
and all other particles are in equilibrium. 
Within this approximation, the collision integral 
$C_i$ of the R.H.S. of the Boltzmann equation 
(\ref{Boltzmann_eqn}) can be expressed as 
\begin{equation}
C_i=-\frac{F_i^{eq}\phi_i}{\tau_i(E_i^*)},
\label{RHS_Boltzmann_relax_time}
\end{equation}
where $\tau_i (E_i^*)$ is the relaxation time.
For $F_i^{eq}={F_i}^{eq}_{\rm (approx)}=
{F_i}^{eq}_{\rm (cl)}(1-{F_i}^{eq}_{(cl)})$,
the relaxation time is obtained as
\begin{eqnarray}
\frac{1-{F_i}^{eq}_{\rm (cl)}}{\tau_i(E_i^*)}
&=&\sum_{jkl}\frac{1}{1+\delta_{ij}}
\int d\Gamma_j d\Gamma_k d\Gamma_l  W(ij|kl) {F_j}^{eq}_{\rm (cl)}
(1-{F_k}^{eq}_{\rm (cl)})(1-{F_l}^{eq}_{\rm (cl)})
\nonumber \\ 
&+&\sum_{kl}\int d\Gamma_k d\Gamma_l W(i|kl)
(1-{F_k}^{eq}_{\rm (cl)})(1-{F_l}^{eq}_{\rm (cl)})\nonumber \\
&+&\sum_{jk}\int d\Gamma_j d\Gamma_k  W(k|ij) {F_j}^{eq}_{\rm (cl)}
(1-{F_k}^{eq}_{\rm (cl)}),
\label{relax_time}
\end{eqnarray}
where $d\Gamma_i =\gamma_i \frac{d^3{\bf {p}}^*_i}{(2\pi)^3}$,
with $\gamma_i$=2 as the spin degeneracy factor for the $i$-th 
nucleon (antinucleon). 
In equation (\ref{relax_time}), the expressions
of $W(ij|kl)$, $W(i|kl)$ and $W(k|ij)$ correspond to the
spin averaged matrix elements for the $2\rightarrow 2$, 
$2\rightarrow 1$ and $1\rightarrow 2$ processes.
This leads to the form of the Boltzmann equation to be given as
\begin{equation} 
\frac{d{F_i({\bf x}, {\bf p}^*_i, t)}}{dt}
=-\frac{F_i^{eq}\phi_i}{\tau_i(E_i^*)}.
\label{Boltzmann_eqn_relaxation_time}
\end{equation}
The L.H.S. of the Boltzmann 
equation, 
$\frac{d{F_i({\bf x}, {\bf p}^*_i, t)}}{dt}
(=\frac{\partial F_i}{\partial t}
+\frac{d {x^k}}{dt} \frac{\partial F_i}{\partial {x^k}} 
+\frac{d {{p_i^*}^k}}{dt} \frac{\partial F_i}{\partial {{p_i^*}^k}})$,
is evaluated by using the local equilibrium form 
of the distribution function, $F_i^{eq}({\bf x}, {\bf p}^*_i,t)$
\cite{Albright_Kapusta_PRC93_014903_2016,Chakraborty_Kapusta_PRC83_014906_2011}, which acts as a source term 
for the collision term on the R.H.S. of the Boltzmann equation.
Equating the tensor structures on both sides
of the above equation, the particular solutions for the 
functions ${\tilde A}_i$,  ${\tilde B}_i$,  ${\tilde C}_i$ 
of the function $\phi_i(x,p)$ given by equation (\ref{phi_i})
are obtained. For a given solution for the function ${\tilde B}_i$, 
$({\tilde B}_i-b)$ is also a solution,
where $b$ is a constant independent of the particle species. 
This ambiguity is resolved by enforcing the Landau-Lifshitz
condition, which requires $\delta T^{0k}=0$ in the local rest 
frame.
The expressions for the coefficients of shear viscosity 
and thermal conductivity
are then obtained as \cite{Albright_Kapusta_PRC93_014903_2016}
\begin{equation}
\eta=\frac{1}{15T}\sum_{i=p,n} 
\int  d\Gamma_i
\frac{{|{\bf p}^*_i|}^4}{{E_i^*({{\bf p}^*_i})}^2}
\tau_i (E_i^*({{\bf p}^*_i})) 
\Big ({f_i}^{eq}({\bf p}^*_i)
+{{\bar f}_i}^{eq}({\bf p}^*_i)\Big)
\label{eta}
\end{equation}
and 
\begin{eqnarray}
\kappa & = &\frac{1}{3}\Big (\frac {\cal W}{\rho_B T}\Big)^2
\sum_{i=p,n} \int d\Gamma_i 
\frac{|{\bf p}^*_i|^2}{{E_i^*({\bf p}^*_i)}^2}\tau_i (E_i^*({\bf p}^*_i)) 
\Big[
\Big (1 -\frac{\rho_B}{\cal W}
\big( E^*_i({\bf p}^*_i)+
(g_{\omega i} \omega +g_{\rho i} \rho)\big)
\Big)^2 
{f_i}^{eq}({\bf p}^*_i)
\nonumber \\
&+&\Big (1+ \frac{\rho_B }{\cal W}
\big( E^*_i({\bf p}^*_i)-
(g_{\omega i} \omega +g_{\rho i} \rho)\big)
\Big)^2 
{{\bar f}_i^{eq}}({\bf p}^*_i)
\Big].
\label{kappa}
\end{eqnarray}
In the above equations, $f_i^{eq}$ and ${\bar f}_i^{eq}$
are the distribution functions of the nucleon 
and antinucleon, given by equation (\ref{quantum_distr_approx}),
which reduce to the classical form given by equation 
(\ref{classical_distr}), in the limit $F_i^{eq}<< 1$. 
In this approximation,
the expressions of the relaxation time, coefficients
of shear viscosity and thermal conductivity, given by
equations (\ref{relax_time}), (\ref{eta}) and (\ref{kappa}),  
reduce to the expressions obtained in the classical approximation
\cite{Albright_Kapusta_PRC93_014903_2016}.
In the present work, we retain the form given by 
equation (\ref{quantum_distr}) to compute the coefficients
of the shear viscosity and thermal conductivity 
\cite{Albright_Kapusta_PRC93_014903_2016}.
In local rest frame, the forms of the distribution functions
for the nucleon and antinucleon
are as given by equation (\ref{equil_distr_fns}) in terms 
of the effective energy 
$E_i^*({\bf p}^*_i)=({{\bf p}^*_i}^2+{m_i^*}^2)^{1/2}$,
with $m_i^*$ as the effective mass (given by equation (\ref{ameff}))
and the effective chemical potential, $\mu_i^*=\mu_i-g_{\omega i} \omega
-g_{\rho i} \rho$. These are computed for given 
density, temperature and isospin asymmetry parameter
from the mean scalar ($\sigma$, $\zeta$ and $\delta$) 
and vector ($\omega$ and $\rho$) fields calculated
in the chiral SU(3) model.
The transport coefficients thus account for
the medium modifications of the nucleons calculated
within the chiral SU(3) model as described 
in the previous section.

As can be seen in the expressions for the transport coefficients,
the relaxation time, $\tau_i$ given by equation (\ref{relax_time}),
is energy dependent, which depends
on the integration variable ${\bf p}^*_i$.
For the system of nucleons as in the present study, 
we make a further assumption and replace the  
momentum-dependent relaxation time, $\tau_i$, 
by a medium-dependent mean value, which is then taken 
outside of the integral \cite{Gavin_NPA435_1985_826}.
For a system of particles with very different relaxation 
times, e.g. for nucleons and pions, however, such an assumption 
would not have been a valid assumption. 
In the dilute gas approximation, the relaxation time
is the same as the collision time \cite{Prakash_Phys_Rep_227_321_1993}.
In the present work, the temperature- and density-dependent
relaxation time of the $i$-th nucleon ($i$=p,n), assumed to be same 
as its collision time, is calculated from its average velocity,
$\langle v_i \rangle$ and the mean free path, $\lambda_i$ using
\begin{equation}
\tau_i=\lambda_i/{\langle v_i \rangle},
\label {tau_i}
\end{equation}
where, 
$\langle v_i\rangle 
\equiv \langle \frac{|{\bf p}^*_i|}{E_i^*({\bf p}^*_i)} \rangle$
and $\lambda_i$ are calculated  
using the expressions
\begin{equation}
\langle v_i\rangle 
= \frac{1}{n_i^{tot}} \int d \Gamma_i
\frac{|{\bf p}^*_i|}{E_i^*({\bf p}^*_i)} 
\Big [f_i^{eq}({\bf p}^*_i)+{\bar f}^{eq}({\bf p}^*_i)
\Big]\;\; {\rm and}\;\; \lambda_i =1/(n_i^{tot} \sigma_{NN}), 
\label {v_almb_i}
\end{equation}
with $n_i^{tot}=\int d \Gamma_i 
\big (f_i^{eq}({\bf p}^*_i)+{\bar f}^{eq}({\bf p}^*_i)\big)$.
In the above, $\sigma_{NN}$ is the total nucleon-nucleon 
cross section, taken to be 40 mb 
\cite{Danielewicz_PLB_146_168_1984,Itakura_PRD77_014014_2008} 
in the present work.
It might be noted here that the study of the effects 
of the medium modifications of the nucleon-nucleon cross sections 
is beyond the scope of the model and have not been taken into 
consideration in the present work.

The relaxation times obtained for the proton and neutron
are used for computing the coefficients of
shear viscosity and thermal conductivity of hot nuclear 
matter. In asymmetric nuclear matter with a nonzero 
value of the asymmetry parameter, $\eta_N$,
the relaxation times for the proton and neutron are different, 
due to the difference in their effective masses and effective chemical
potentials. The isospin asymmetry of the medium is observed
to lead to higher values of $\eta/s$ and $\kappa$, 
and the isospin asymmetry effect on the thermal conductivity
is observed to be quite appreciable at high densities,
as shall be discussed in the next section. 

\begin{figure}
\vskip -2.in
\includegraphics[width=14cm,height=16cm]{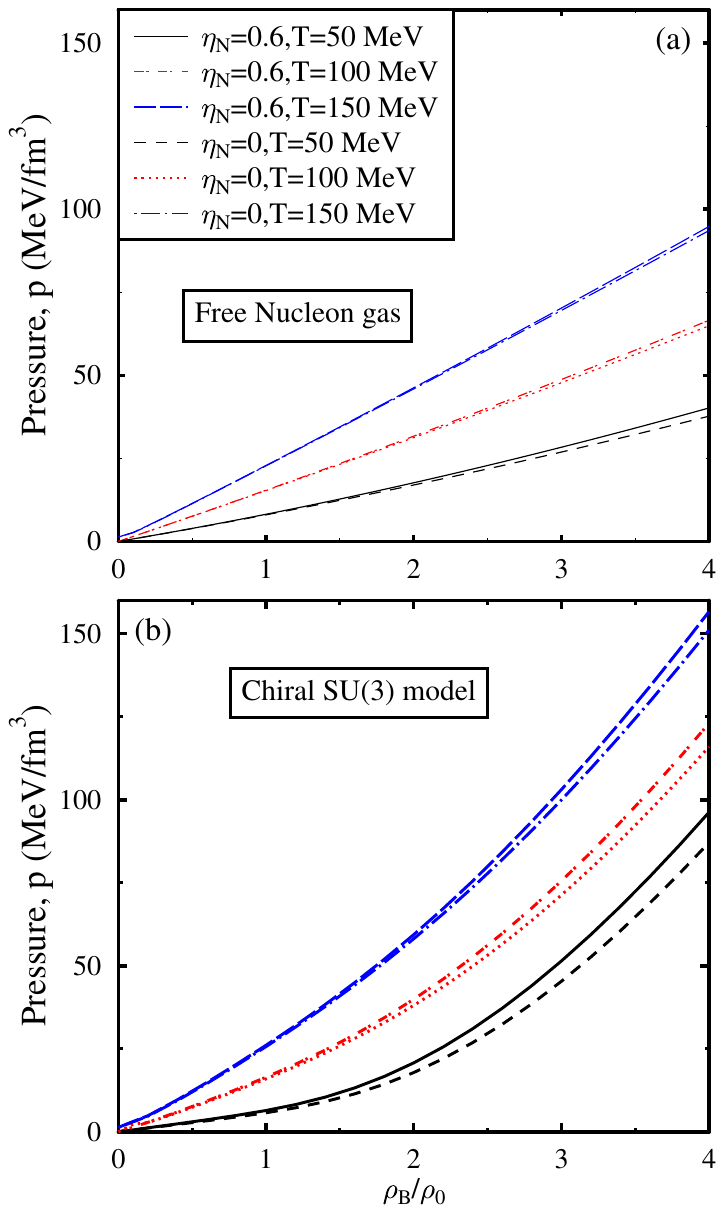} 
\vskip -0.3in
 \caption{Pressure, p (in MeV/fm$^3$)
is plotted as a function of $\rho_B/\rho_0$, the baryon density in units 
of nuclear matter saturation density, 
for different values of temperature, for isospin symmetric 
($\eta_N$=0) and asymmetric nuclear matter (with asymmetry parameter, 
$\eta_N$=0.6). The effects due to the medium modified
nucleons calculated using the chiral SU(3) model
(shown in subplot (b)) are compared with the case of free nucleon gas
(shown in subplot (a)).
}
\label{Dens_Pressure}
 \end{figure}

\begin{figure}
\vskip -2.in
\includegraphics[width=14cm,height=16cm]{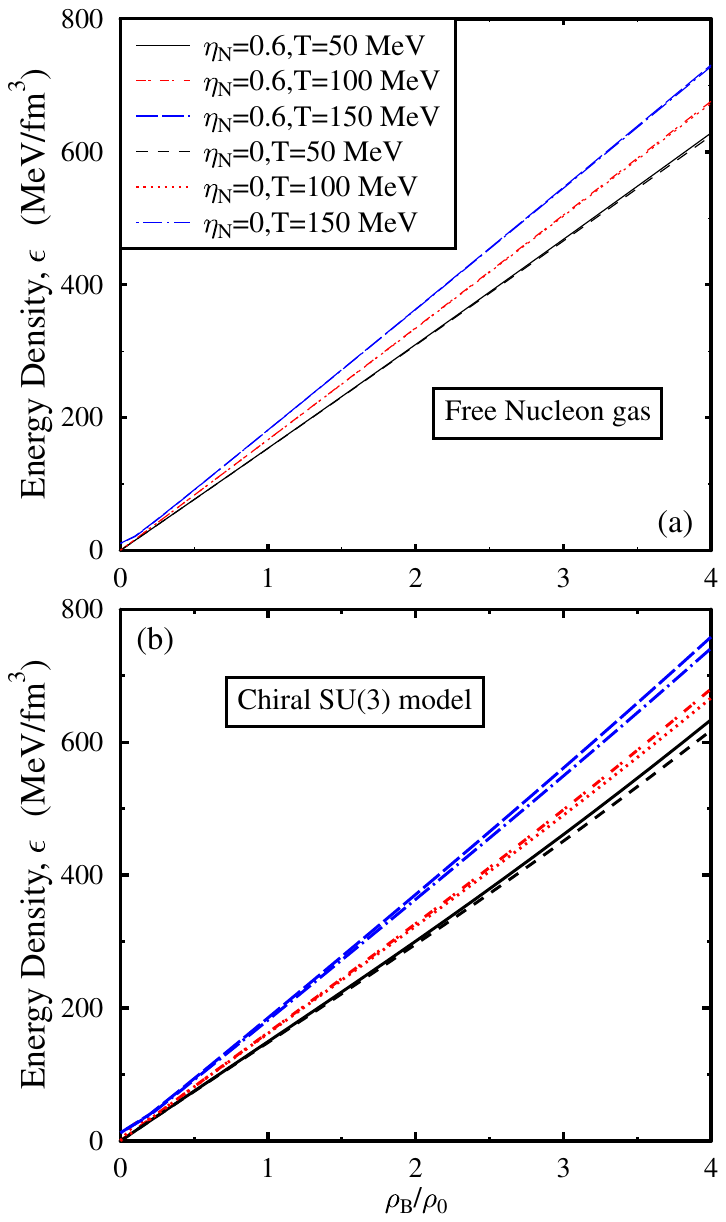} 
\vskip -0.3in
 \caption{Energy density (in MeV/fm$^3$) is plotted 
as a function of $\rho_B/\rho_0$,
the baryon density in units of nuclear matter saturation density, 
for different values of temperature, for isospin symmetric 
($\eta_N$=0) and asymmetric nuclear matter (with asymmetry parameter, 
$\eta_N$=0.6). The effects due to the medium modified
nucleons calculated using the chiral SU(3) model
(shown in subplot (b)) are compared with the case of free nucleon gas
(shown in subplot (a)).
}
\label{Dens_Energy_Density}
 \end{figure}

\begin{figure}
\vskip -2.in
\includegraphics[width=14cm,height=16cm]{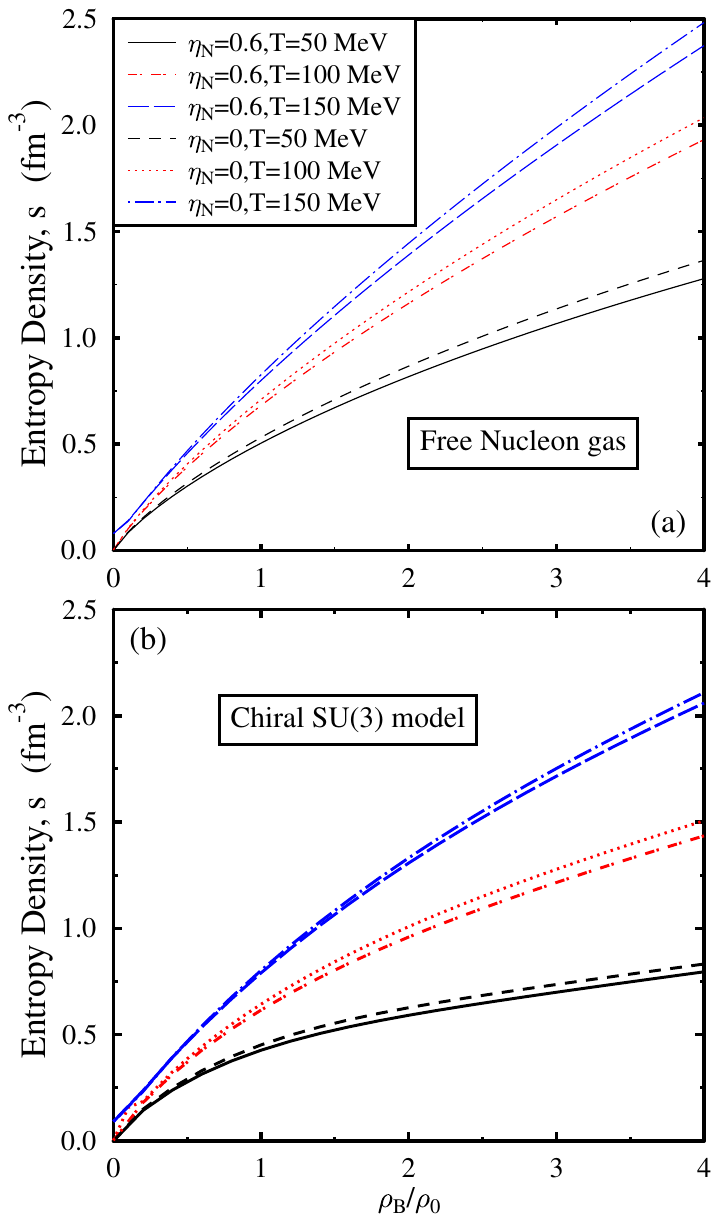} 
\vskip -0.3in
 \caption{Entropy density, $s$ (in fm$^{-3}$) 
is plotted as a function of $\rho_B/\rho_0$,
the baryon density in units of nuclear matter saturation density, 
for different values of temperature, for isospin symmetric 
($\eta_N$=0) and asymmetric nuclear matter (with asymmetry parameter, 
$\eta_N$=0.6). The effects due to the medium modified
nucleons calculated using the chiral SU(3) model
(shown in subplot (b)) are compared with the case of free nucleon gas
(shown in subplot (a)).
}
\label{Dens_Entr}
 \end{figure}

\begin{figure}
\vskip -2.in
\includegraphics[width=14cm,height=16cm]{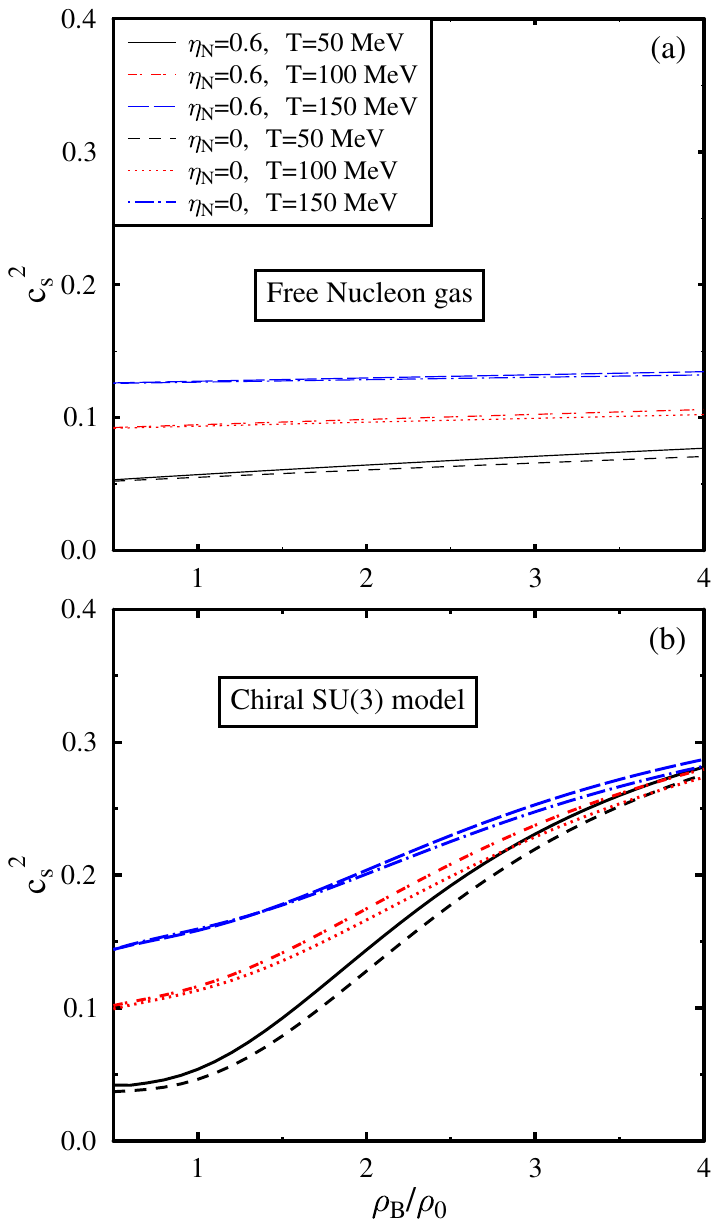} 
\vskip -0.3in
 \caption{$c_s^2$ is plotted as a function of $\rho_B/\rho_0$,
the baryon density in units of nuclear matter saturation density, 
for different values of temperature, for isospin symmetric 
($\eta_N$=0) and asymmetric nuclear matter (with asymmetry parameter, 
$\eta_N$=0.6). The effects due to the medium modified
nucleons calculated using the chiral SU(3) model
(shown in subplot (b)) are compared with the case of free nucleon gas
(shown in subplot (a)).
}
\label{Dens_cs2}
 \end{figure}

\begin{figure}
\vskip -2.in
\includegraphics[width=14cm,height=16cm]{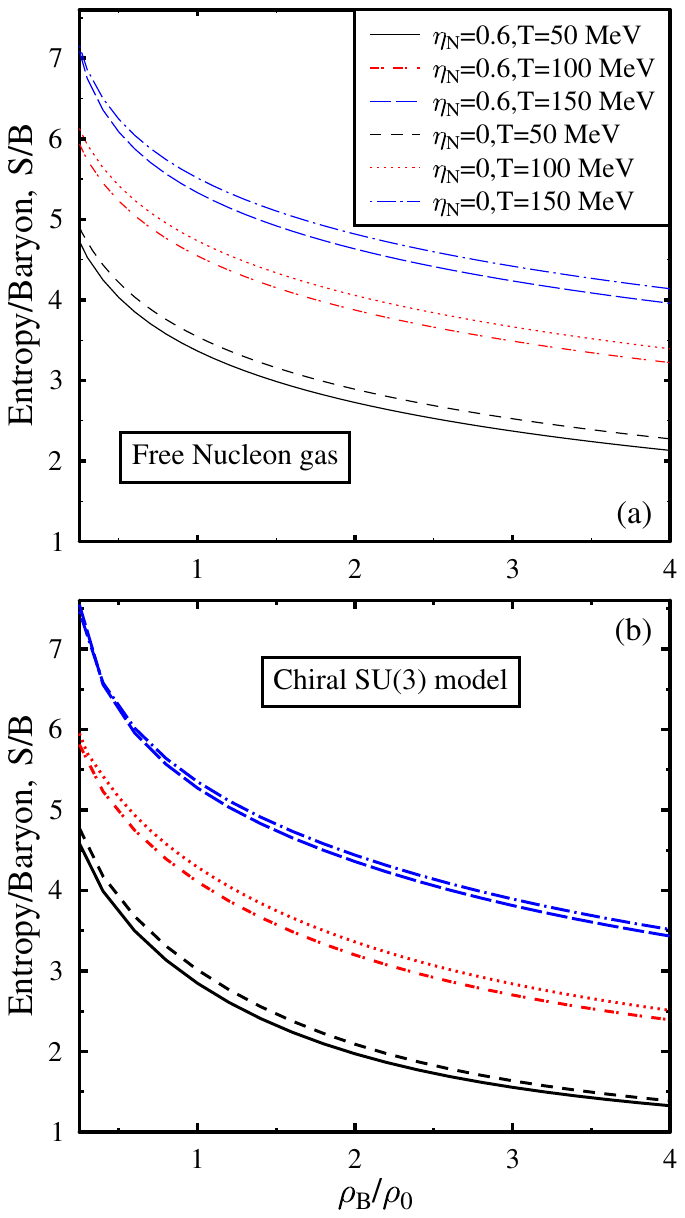} 
\vskip -0.3in
 \caption{Entropy per baryon, $S/B$
is plotted as a function of $\rho_B/\rho_0$,
the baryon density in units of nuclear matter saturation density, 
for different values of temperature, for isospin symmetric 
($\eta_N$=0) and asymmetric nuclear matter (with asymmetry parameter, 
$\eta_N$=0.6). The effects due to the medium modified
nucleons calculated using the chiral SU(3) model
(shown in subplot (b)) are compared with the case of free nucleon gas
(shown in subplot (a)).
}
\label{Dens_Entr_baryon}
 \end{figure}

\begin{figure}
\vskip -2.in
\includegraphics[width=14cm,height=16cm]{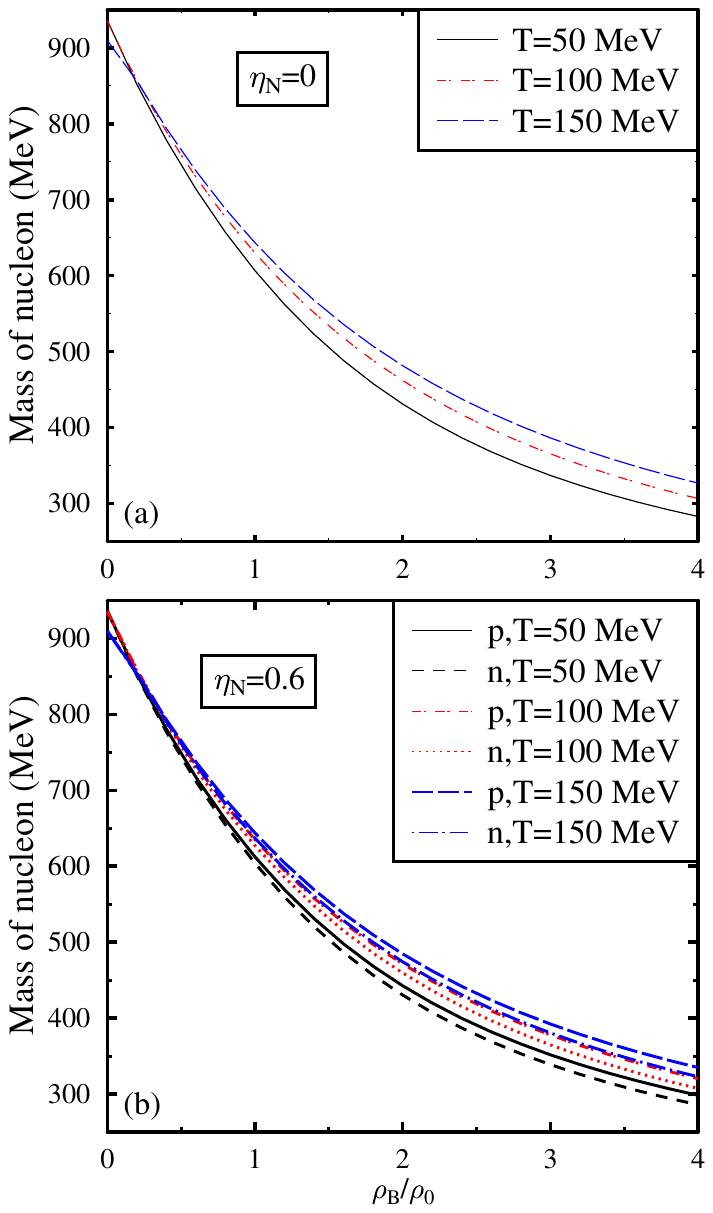} 
\vskip -0.3in
 \caption{Masses of protons and neutrons, calculated within the
chiral SU(3) model, are plotted as functions of 
$\rho_B/\rho_0$,
the baryon density in units of nuclear matter saturation density, 
for different values of temperature, for isospin
symmetric ($\eta_N$=0) and asymmetric (with $\eta_N$=0.6) nuclear matter
in panels (a) and (b) respectively.
}
\label{Dens_mass_p_n}
 \end{figure}

\begin{figure}
\vskip -2.in
\includegraphics[width=14cm,height=16cm]{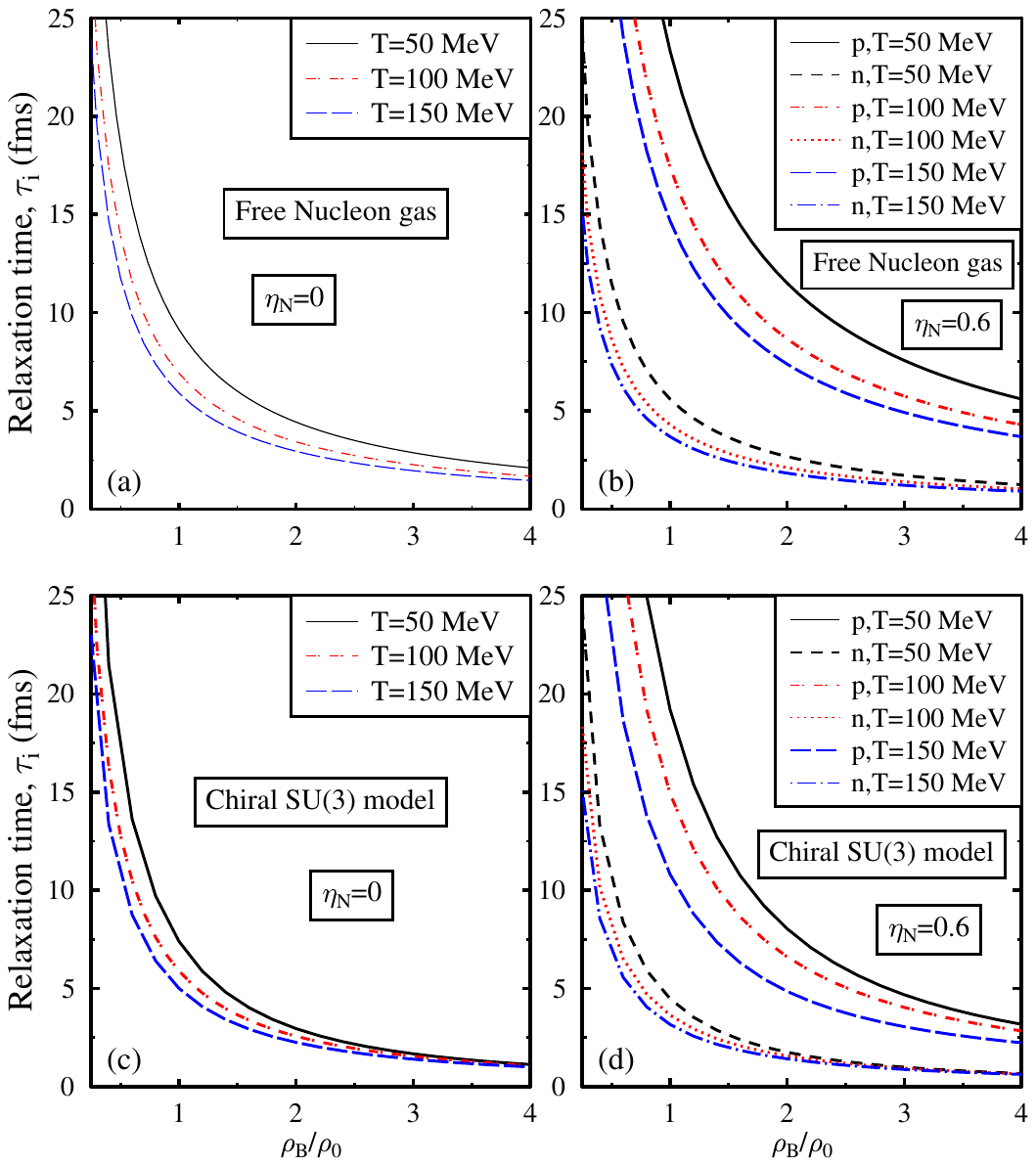} 
\vskip -0.3in
 \caption{Relaxation times of proton and neutron (in Fermi) 
are plotted as functions of $\rho_B/\rho_0$,
the baryon density in units of nuclear matter saturation density, 
for different values of temperature, for isospin
symmetric ($\eta_N$=0) and asymmetric (with $\eta_N$=0.6) nuclear matter
in free nucleon gas in subplots (a) and (b) respectively 
and accounting for the medium modifications of nucleons using
chiral SU(3) model in subplots (c) and (d) respectively. 
}
\label{Dens_tau_p_n}
 \end{figure}

\begin{figure}
\vskip -2.in
\includegraphics[width=14cm,height=16cm]{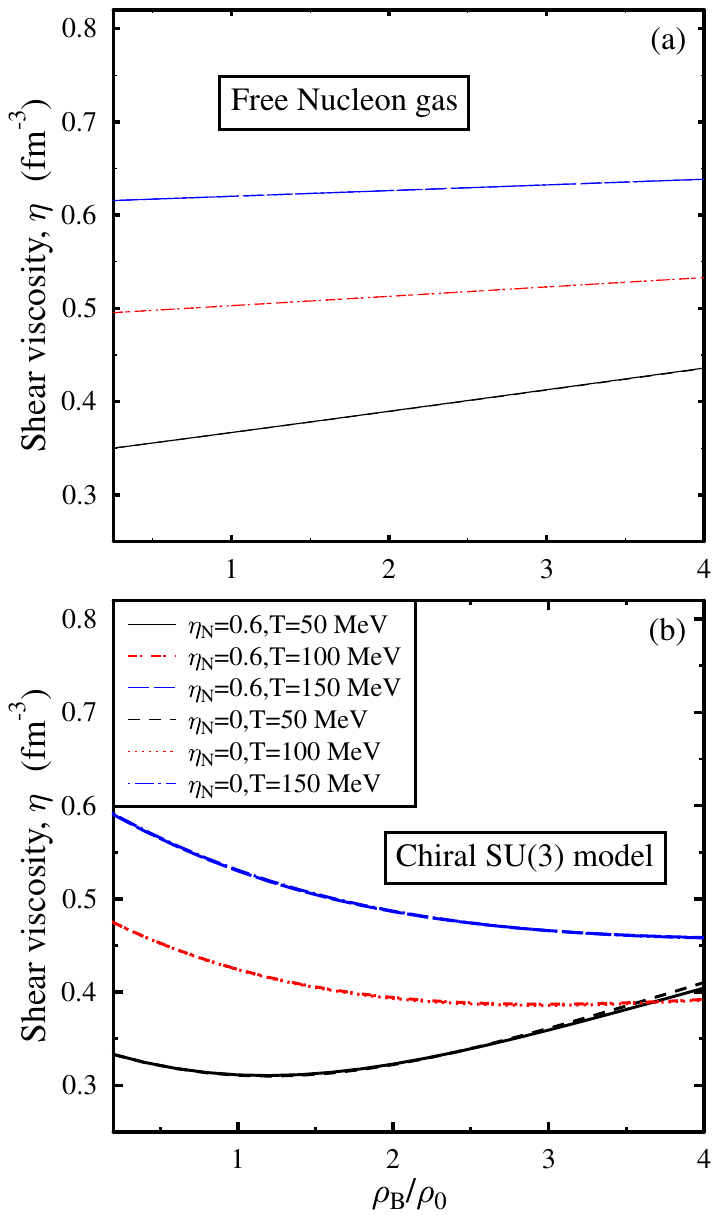} 
\vskip -0.3in
 \caption{Shear viscosity coefficient, $\eta$, is plotted 
as a function of $\rho_B/\rho_0$,
the baryon density in units of nuclear matter saturation density, 
for different values of temperature, for isospin symmetric 
($\eta_N$=0) and asymmetric nuclear matter (with asymmetry parameter, 
$\eta_N$=0.6). 
This is plotted for free nucleon gas (in subplot (a))
and for the chiral SU(3) model with the medium modified 
nucleons (in subplot (b)). 
}
\label{Dens_shear_visc}
 \end{figure}

\begin{figure}
\vskip -2.in
\includegraphics[width=14cm,height=16cm]{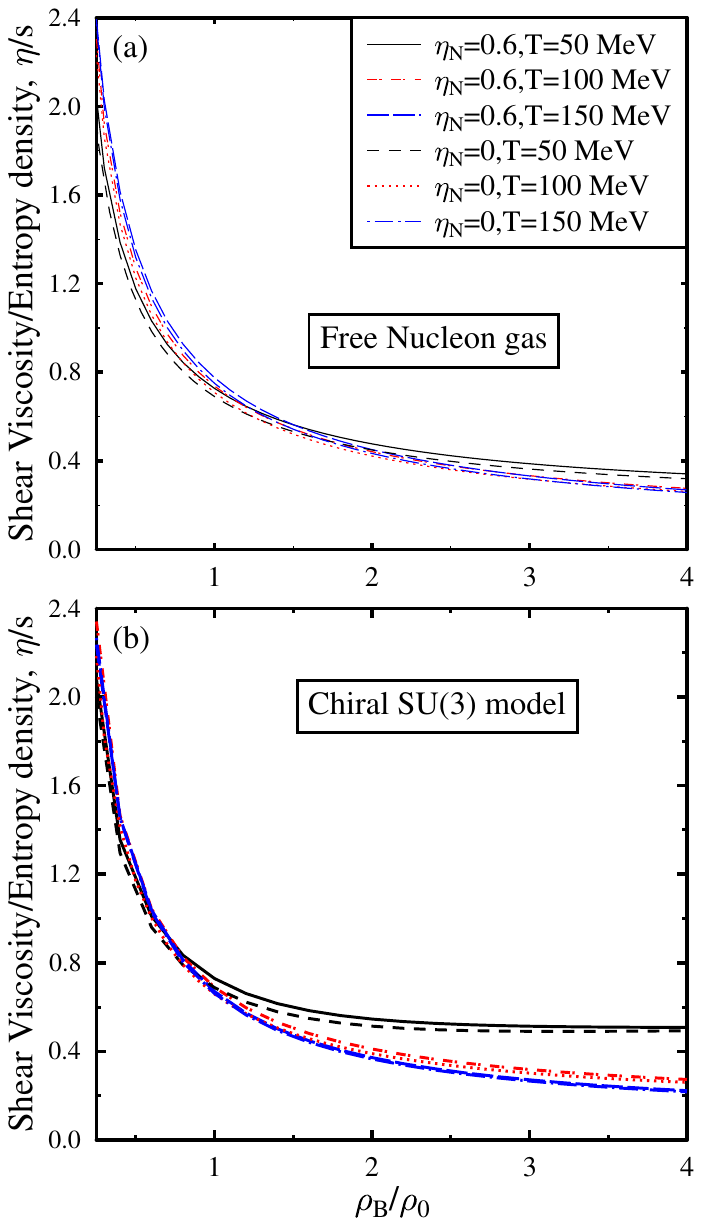} 
\vskip -0.3in
 \caption{Shear viscosity coefficient to entropy density ratio, $\eta/s$,
is plotted as a function of $\rho_B/\rho_0$,
the baryon density in units of nuclear matter saturation density.
for different values of temperature, for isospin symmetric 
($\eta_N$=0) and asymmetric nuclear matter (with asymmetry parameter, 
$\eta_N$=0.6). 
This is plotted for free nucleon gas (in subplot (a))
and for the chiral SU(3) model with the medium modified 
nucleons (in subplot (b)). 
}
\label{Dens_shear_visc_entr}
 \end{figure}

\begin{figure}
\vskip -2.in
\includegraphics[width=14cm,height=16cm]{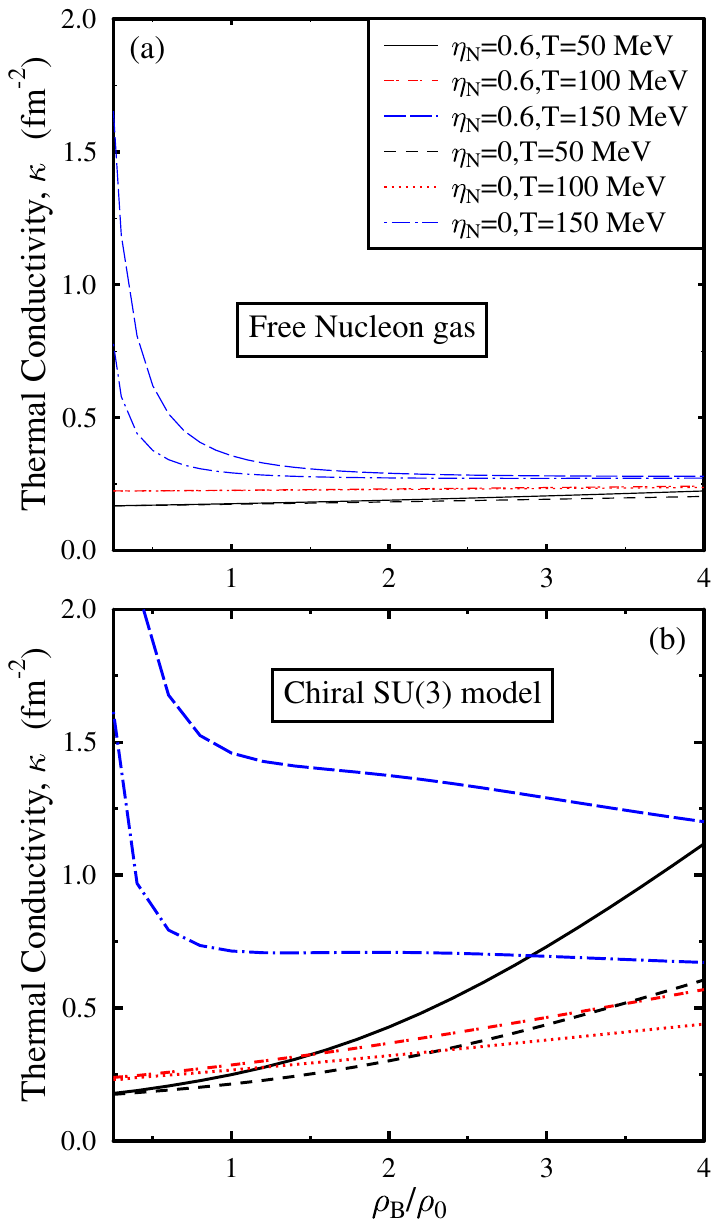} 
\vskip -0.3in
 \caption{Thermal conductivity coefficient, $\kappa$ is plotted 
as a function of $\rho_B/\rho_0$,
the baryon density in units of nuclear matter saturation density, 
for different values of temperature, for isospin symmetric 
($\eta_N$=0) and asymmetric nuclear matter (with asymmetry parameter, 
$\eta_N$=0.6). This is plotted for free nucleon gas (in subplot (a))
and for the chiral SU(3) model with the medium modified 
nucleons (in subplot (b)). 
}
\label{Dens_therm_cond}
 \end{figure}

\section{Results and discussions}
In this section, we discuss the results for the thermodynamic 
and transport properties of hot asymmetric nuclear matter
studied in the present work using a chiral SU(3)
model. The shear viscosity, $\eta$ and thermal conductivity ($\kappa$) 
are computed using the in-medium properties of the nucleons. 
For given values of the baryon density, $\rho_B$,
temperature, $T$ and the isospin asymmetry parameter,
$\eta_N$, these are calculated using equations (\ref{eta}) and
(\ref{kappa}), from the equilibrium distribution functions 
of the nucleons obtained from their effective masses and 
effective chemical potentials calculated within the model.
The thermodynamic quantities and transport 
coefficients, incorporating the effects due to the medium modifications 
of the nucleons using the chiral SU(3) model,
are compared with the results obtained for the
free nucleon gas.
The values of the parameters of the chiral SU(3) model
used in the present work are as follows \cite{AM_AK_PRC81_065204_2010}.
The scalar-isoscalar meson-nucleon couplings, $g_{\sigma i}$=10.6 and
$g_{\zeta i}=-0.47$, $i$=p,n, are fitted to the vacuum nucleon mass,
taken to be $m_i^{vac}$=939 MeV.
The scalar-isovector meson-nucleon coupling 
$g_{\delta p}=2.5=-g_{\delta n}$,
and the masses of $\sigma$, $\zeta$ and $\delta$ fields are
466.5, 1024.5 and 899.5 MeV respectively.
The vector meson-nucleon couplings are $g_{\omega i}=13.3$,
$g_{\rho p}=-g_{\rho n}=5.5$, the strength of the self-interactions
of the vector fields, $g_4$=79.9 and the masses of the vector fields
are $m_\omega$=780.6 and $m_\rho$=761 MeV. 
The parameters of the scalar potential
are taken to be $k_0=2.54$, $k_1=1.35$, $k_2=-4.78$, 
$k_3=-2.77$, $k_4=-0.22$ and d=0.064. 
The vacuum values of the scalar-isoscalar fields, $\sigma$ and
$\zeta$ and the dilaton field, $\chi$ are $-93.3$, $-106.6$
and 409.8 MeV respectively.

In figures \ref{Dens_Pressure}, \ref{Dens_Energy_Density}
and \ref{Dens_Entr}, the density dependence of the 
thermodynamic quantities, the pressure, the energy density 
and the entropy density are shown. These are plotted 
as functions of the baryon density in units of nuclear matter
saturation density, $\rho_B/\rho_0$, for different temperatures
for isospin symmetric ($\eta_N$=0) and asymmetric 
(with isospin asymmetry parameter, 
$\eta_N(=(\rho_n-\rho_p)/\rho_B)$=0.6) 
nuclear matter.
The results of the chiral SU(3) model, accounting for the
in-medium effects are compared with the results
for the free nucleon gas. The medium effects 
are small for low densities, yielding similar values
for the chiral model and the free nucleon gas. 
These are observed to be larger when the density is increased,
since the medium modifications of the nucleons and 
the mean fields become larger at higher densities.
At high densities, the increase in the pressure in the chiral SU(3) model 
is observed to be appreciably larger as compared to for the 
free nucleon gas.
For example, for $\rho_B=4 \rho_0$, for symmetric
nuclear matter, the values of the pressure 
are observed to be around 87, 116 and 151
MeV/fm$^3$ as compared to the values of 
38, 65 and 94 MeV/fm$^3$
for free nucleon gas, for temperature, 
T=50, 100 and 150 MeV respectively. 
The appreciably larger values of the pressure 
in the chiral model as compared to free nucleon gas
at high densities can be 
understood as follows. For the free nucleon gas, the 
pressure has contribution only from the last term of the thermodynamic
potential given by equation (\ref{thermo_total}), with the vacuum mass
of the nucleon, $m_i^{vac}$ and the chemical potential, 
$\mu_i^{(fr)}$ obtained from the $i$-th nucleon density, $\rho_i$, 
$i$=p,n, for given baryon density, isospin asymmetry and temperature.  
For the chiral model, this term is modified due to the effective 
nucleon masses and chemical potentials, and since there is
appreciable decrease in the nucleon mass at high densities,
the value of this term is observed to be modified from the 
value of around 38 MeV/fm$^3$
for free nucleon gas to the value of 54 MeV/fm$^3$ in the 
chiral SU(3) model at $\rho_B=4\rho_0$ for symmetric nuclear 
matter at T=50 MeV. 
Furthermore, there is substantial contribution (of around
33 MeV/fm$^3$) from the other terms of the thermodynamic potential 
arising from the mean scalar and vector fields, giving the 
value for the total pressure to be 87 MeV/fm$^3$.
In the presence
of isospin asymmetry in the medium, the contributions
from the proton and neutron are different (of around 8 and 50.5
MeV/fm$^3$
respectively) from the last term of the thermodynamic potential,
and, 37.5 MeV/fm$^3$
from the other terms arising from the mean fields, 
yielding a larger value
of around 96 
MeV/fm$^3$
for the pressure in the isospin asymmetric medium
as compared to the value of 87 
MeV/fm$^3$
for symmetric nuclear matter
at T=50 MeV.
 
In figure \ref{Dens_Energy_Density}, 
the energy density (in MeV/fm$^3$) is plotted 
as a function of the baryon density for given temperatures
and isospin asymmetry parameters
for the chiral model and 
compared with the values of 
the free nucleon gas. For the higher value of the baryon density, 
$\rho_B=4\rho_0$ and temperatures, T=50, 100 and 150 MeV,
the values of the energy density 
are observed to be 
618, 667 and 741
MeV/fm$^3$ 
in the chiral SU(3) model and 
624, 673 and 728
MeV/fm$^3$ 
for the free nucleon gas.
It might be noted here that for the free nucleon gas, the
energy density has contribution only from the first term
of equation (\ref{energy_density}) with the nucleons
with their vacuum masses $m_i^{vac}$, and 
$\mu_i^{(fr)}$ calculated from the nucleon densities $\rho_i$.
The contribution of this term in the chiral SU(3) model
is smaller than the free nucleon gas at high densities
due to much smaller mass of the nucleons. However,
the energy density in the chiral SU(3) model gets appreciable 
contributions from the other terms
of equation (\ref{energy_density}). This is observed as the values 
of the energy densities to be quite similar in their values
in the two models.
For example, for the isospin symmetric nuclear matter
for $\rho_B=4\rho_0$ and T=50 MeV, the contribution 
to the energy density 
from the first term is
624 and 271 
MeV/fm$^3$
for the free nucleon gas and the chiral SU(3) model,
the latter, however, has a contribution of 347 
MeV/fm$^3$
from the other terms,
yielding the values of the energy density to be 624  
and 618 
MeV/fm$^3$
in the free nucleon gas and the chiral SU(3) model
respectively.

As can be seen from figure \ref{Dens_Entr}, 
the interactions of the nucleons are observed 
to lead to a noticeable drop in the value of entropy density 
at high densities due to the much larger drop
in the nucleon mass in the nucleon mass,
in addition to the modification due to the 
effective chemical potential, which modifies the distribution
functions in the chiral SU(3) model in the presence of
interactions with the mean scalar and vector fields.
The values of the entropy density (in fm$^{-3}$)
are observed to be 0.83 (0.8), 1.51 (1.44) 
and 2.11 (2.06) in the chiral SU(3) model, 
as compared to 1.36 (1.28), 2.04 (1.93) and 2.48 (2.37), 
in the free nucleon gas
for T=50, 100 and 150 MeV, for $\rho_B=4 \rho_0$ in isospin 
symmetric (asymmetric matter with value of the
asymmetry parameter, $\eta_N$=0.6) nuclear matter.
The isospin asymmetry is observed to raise the values of
the pressure and the energy density, whereas for the entropy density,
the isospin asymmetry leads to a smaller value, both for the
free nucleon gas and in the chiral SU(3) model.

Figure \ref{Dens_cs2} shows the density dependence of the square
of the speed of sound $c_s^2 (={\partial p}/{\partial \epsilon})$
for temperatures T=50, 100 and 150 MeV respectively,
both for symmetric and asymmetric nuclear matter.
The effects due to medium
modifications of the nucleons calculated in  
the chiral SU(3) model are shown in subplot (b), 
and, are compared with the results for the free nucleon gas 
shown in subplot (a) of the same figure.
For the free nucleon gas, due to the (almost) linear 
rise in the pressure (as well as energy density)
with density, as observed in figures \ref{Dens_Pressure} 
and \ref{Dens_Energy_Density}, the value 
of $c_s^2$ is observed to have a slow increase
for T=50 MeV (from the value of 0.052 at $0.5\rho_0$ to 
0.07 at 4$\rho_0$), whereas, it remains almost constant
for the higher values of the temperatures.  
In the chiral SU(3) model, however, 
there is observed to be an increase in the value of
$c_s^2$ with density, which is due to the much larger
nonlinear increase of the pressure as compared 
to the (almost) linear increase of energy density. 
The isospin asymmetry is observed to lead to higher
values of $c_s^2$, both for free nucleon gas and
in chiral SU(3) model with medium modified nucleons.
At high densities, in the chiral SU(3) model,
it is observed that the values of $c_s^2$ are very similar 
for different values of the temperature, unlike the case
of the free nucleon gas.
for example, the values of $c_s^2$ for density, 
$\rho_B=4\rho_0$, are 0.276 (0.281), 0.274 (0.28) 
and 0.28 (0.287) for symmetric nuclear matter 
(asymmetric with $\eta_N$=0.6) 
for values of temperature T=50, 100 
and 150 MeV, respectively, in the chiral SU(3) model, which may be 
compared with the values of 0.07 (0.077), 0.102 (0.106)
and 0.132 (0.135) for free nucleon gas, 
for the same temperatures and baryon density. 

In Fig. \ref{Dens_Entr_baryon}, we show the density dependence
of the ratio of entropy density to baryon density ($\equiv$ S/B) 
for the temperatures T=50, 100 and 150 MeV both for free nucleon
gas and chiral SU(3) model with medium modified nucleons.
There is observed to be an increase in the entropy per baryon, S/B  
as the temperature is raised. The effect of isospin asymmetry 
is observed to lead to a smaller value of $S/B$ as compared to
isospin symmetric matter at the same density, 
similar to the isospin asymmetry effect on the entropy density
(see figure \ref{Dens_Entr}), as expected.
The trend of the dependence on the density for $S/B$
is similar for the free nucleon gas and for the chiral SU(3) model
including medium effects of the nucleons. 
However, the value of $S/B$ is observed to be lower 
in the chiral SU(3) model 
as compared to the free nucleon gas,
for the same densities, which is due to the smaller values of
the entropy density in the presence of medium effects 
in the chiral SU(3) model.

In Fig.\ref{Dens_mass_p_n},
the effective masses of the proton and neutron 
(given by equation (\ref{ameff})) are plotted 
as functions of the baryon density (in units of 
nuclear matter saturation density)
both for symmetric and asymmetric nuclear matter 
for values of temperature T=50, 100 and 150 MeV.
For the symmetric nuclear
matter ($\eta_N$=0), the masses of the proton and neutron
are degenerate. These masses become nondegenerate in the presence 
of isospin asymmetry due to the interaction with the 
isovector-scalar $\delta$ meson, which leads to a higher 
mass of the proton as compared to the neutron. There is observed 
to be a decrease in the masses of the nucleons in the nuclear matter
with increase in baryon density for a given
temperature, and, a larger value of the temperature 
is observed to lead to a smaller drop in the mass of the nucleon.
This behavior of the nucleon mass with
temperature has also been observed within the
framework of quantum hadrodynamics (QHD) \cite{Li_Ko}.
However, at the higher temperature of T=150 MeV, 
one observes the nucleon mass to be different (and smaller) 
than the vacuum value for density $\rho_B$=0, which 
arises due to the nonzero fluctuations
of the scalar fields from their vacuum values. 
This behavior of the nucleon mass to be different
from the vacuum value at zero density 
and large finite temperatures, has already been
discussed in the literature within QHD
\cite{Furnstahl_Serot} as well as in the chiral SU(3) model 
\cite{kristof1}.
It might be noted here that, for zero density, the nonzero 
deviations of the values of the scalar fields from their vacuum values 
(leading to the nucleon masses to be different from their
vacuum values) is observed for temperature T=150 MeV,
which is close to the value of the pseudocritical temperature
predicted from lattice QCD and the validity of hadronic description 
may be questionable around this temperature.
In asymmetric nuclear matter, when the densities of the proton and
neutron are different, the effective masses and the
effective chemical potentials of the proton and neutron are
no longer degenerate as in symmetric nuclear matter.
This leads to the relaxation times of the proton and neutron, 
which are calculated using equations (\ref{tau_i})--(\ref{v_almb_i}), 
to be different. 
The relaxation times of the nucleons are plotted as functions
of the baryon density (in units of nuclear matter saturation 
density) for different temperatures in figure \ref{Dens_tau_p_n}.
These are plotted for isospin symmetric ($\eta_N$=0) and 
asymmetric (with $\eta_N$=0.6) free nucleon gas 
in subplots (a) 
and (b) respectively and accounting for the medium modifications 
of nucleons within chiral SU(3) model in subplots (c) and (d) 
respectively. 
There is observed to be a drop with increase in the
baryon density for the temperatures T=50, 100 and 150 MeV
considered in the present study. 
The relaxation times are observed to
be higher as the temperature is lowered. 
The isospin asymmetry is observed to lead to a difference
in the relaxation times of the proton and the neutron,
for the free nucleon gas as well as in the chiral SU(3) model,
as can be seen from figure \ref{Dens_tau_p_n}.
In the presence of isospin asymmetry, the proton has a
larger relaxation time as compared to the neutron, 
due to smaller value of its (total) density, which leads to
a larger value of the mean free path (see equation 
(\ref{tau_i})). 
In the chiral SU(3) model, smaller masses of the nucleons 
as compared to the free nucleon gas with the vacuum masses,
leads to the average velocity of the nucleons 
(see equation (\ref{v_almb_i})) to be larger, leading to smaller
values of the relaxation times of the nucleons
for the chiral SU(3) model as compared to the free nucleon gas.
In symmetric nuclear matter, the values of nucleon relaxation 
time (in fermis) of 7.4 (1.13), 5.9 (1.08) and 5.01 (1) 
for T=50, 100 and 150 MeV respectively, at $\rho_B=\rho_0 
(4\rho_0)$, obtained using the chiral SU(3) model,
may be compared to the values of 9.14 (2.1), 6.9 (1.68) 
and 5.9 (1.46) for the free nucleon gas.
For T=100 MeV, the values of 5.9 (1.08) at $\rho_0 (4\rho_0)$ 
in the chiral SU(3) model of the present work 
may be compared with the values of 
around 4.2 (2) 
in Ref. \cite{Danielewicz_PLB_146_168_1984} 
and of around 2 (0.4) of Ref. 
\cite{NPA573_554_1994_Monras_Trans_coeffs_Nucl_Neutron_matter}.
In the asymmetric nuclear matter, the relaxation time 
is observed to be higher for the proton as compared
to the neutron, mainly due to smaller number density
of the proton as compared to neutrons.
For T=50, 100 and 150 MeV, the values of the relaxation
time (in fermis) for the proton (neutron)
are observed to be 19.2 (4.48), 14.92 (3.65) and 10.82 (3.17)
for density $\rho_0$, which are modified to 
3.18 (0.68), 2.85 (0.66) and 2.24 (0.63) at
the higher density of $\rho_B=4\rho_0$.
The isospin asymmetry of the nuclear medium 
is thus observed to lead to large difference 
in the relaxation times of the proton and neutron. 

The shear viscosity coefficient, $\eta$ (in units of ${\rm fm}^{-3}$) 
is plotted as a 
function of $\rho_B/\rho_0$ in Fig. \ref{Dens_shear_visc}
for T=50, 100 and 150 MeV, both for symmetric ($\eta_N$=0) 
and asymmetric (with $\eta_N$=0.6) nuclear matter. 
This is plotted both for free nucleon gas (in panel (a)) 
as well as with medium modified nucleons calculated
in the chiral SU(3) model (in panel (b)). 
In the free nucleon gas, the value of 
$\eta$ is observed to be larger as temperature
is raised and there is a steady increase 
with increase in density. 
However, in the presence of medium effects
on the nucleons using the chiral SU(3) model, the behavior
of $\eta$ is observed to be different as compared 
to the free nucleon gas. 
For temperature, T=50 MeV, there is observed to be an 
initial drop of $\eta$ with density up to around $\rho_0$,
which is observed to increase as the density is further raised.
The density dependence of the shear viscosity 
$\eta$ at a particular value of the temperature, as can be 
seen from the expression given by equation (\ref{eta}),
arises due to the density dependence of the relaxation 
time $\tau_i$ (calculated using equation (\ref{tau_i}))
and the momentum integral (modulo $\tau_i$).
The former decreases with density whereas the
latter increases with density due to higher contributions
to the integrand from larger values of 3-momenta 
(due to the larger polynomial power)
at higher densities. It is the competition between these
two which gives rise to the observed behavior
of the shear viscosity with density.
The effect of the isospin asymmetry is observed
to lead to a marginally lower value 
of the shear viscosity coefficient.
The higher temperature, T=100 MeV, shows a similar trend,
but the density above which there is observed to be a rise 
is around 2.5 $\rho_0$ and the increase is much slower than 
the case of T=50 MeV. For T=150 MeV, the value of $\eta$
is observed to drop with density and remains almost constant
for densities higher than around $3.5 \rho_0$. 
For T=50 MeV, the rapid increase in the value of $\eta$
with density for densities higher than $\rho_0$ arises due to
appreciable contributions from higher momenta, 
due to the large increase in the value of the difference 
$(\mu_i^*-m^*_i)$ (in MeV) with density,
e.g., from the value of around 75 at $\rho_0$ 
to around 192 at $4\rho_0$ in symmetric nuclear matter.
For the higher values of the temperature,
T=100 and 150 MeV, as the density is increased,
the drop in the magnitude of the relaxation time $\tau_i$
is either similar or marginally larger as compared
to the increase in the magnitude of the integral (modulo $\tau_i$) 
in the expression of the coefficient of shear viscosity 
given by equation (\ref{eta}). This leads to
the value of $\eta$ for high densities
to be slowly decreasing or remaining almost constant 
with increase in density.
The value of the shear viscosity coefficient, $\eta$ 
(in ${\rm fm}^{-3}$) of 
0.42 for T=100 MeV for symmetric nuclear matter 
at $\rho_B=\rho_0$ of the present work 
may be compared 
to the values of 0.3 \cite{Danielewicz_PLB_146_168_1984} and 
0.28 \cite{A_S_Khvorostukhin_NPA845_106_2010},
and, the value of 0.39 at $\rho_B=4\rho_0$
may be compared to the values of around 0.33 
\cite{Danielewicz_PLB_146_168_1984}, 
0.39 \cite{A_S_Khvorostukhin_NPA845_106_2010}
0.15 \cite{NPA573_554_1994_Monras_Trans_coeffs_Nucl_Neutron_matter},
and, to the value of around 0.3 obtained 
in the relaxation time approximation
\cite{Hakim_Mornas_PRC47_2846_1993}.
In the chiral SU(3) model, for asymmetric nuclear matter 
(with $\eta_N$=0.6),
for T=50 MeV, the value of the shear viscosity coefficient, $\eta$ 
is observed to be marginally larger as compared to symmetric nuclear
matter up to $\rho_B\sim 2.3 \rho_0$, whereas, 
$\eta$ is observed to be smaller for higher densities,
with values of $\eta$ (in ${\rm fm}^{-3}$) 
obtained as 0.41 (0.4) for the symmetric (asymmetric)
nuclear matter at $\rho_B=4\rho_0$.  
The effects of isospin asymmetry 
on the value of the shear viscosity coefficient
are observed to be marginal for free nucleon gas
as well as in the chiral model. 

The density dependence of the shear viscosity coefficient 
to the entropy density ratio,
$\eta/s$ is shown in Fig. \ref{Dens_shear_visc_entr},
for hot symmetric as well as asymmetric nuclear matter.
These are plotted both for free nucleon
gas (in panel (a)) as well as with medium modified nucleons calculated
in the chiral SU(3) model (in panel (b)). 
There is observed to be a drop of $\eta/s$ with increase in baryon density
and the value is observed to be slowly (marginally) decreasing
for higher values of the densities, 
similar to the trend observed in the literature
\cite{A_S_Khvorostukhin_NPA845_106_2010,NPA573_554_1994_Monras_Trans_coeffs_Nucl_Neutron_matter,Itakura_PRD77_014014_2008}.
The drop is observed to be larger for higher values of the temperature.
For free nucleon gas, the values of $\eta$ as well as 
$s$ increase with density (see figures \ref{Dens_Entr} and 
\ref{Dens_shear_visc}). However, for the lower temperature, T=50 MeV,
the contribution to $\eta/s$ from $\eta$ ($s$) is dominated 
by the contribution from $s$ ($\eta$) for the low (high) densities, 
which is observed as a larger drop of $\eta/s$
for densities up to around 1.3 $\rho_0$. 
The density dependence of $\eta/s$ is observed to be
opposite for the higher values of the temperature.
In the chiral SU(3) model, at low densities,
the drop of $\eta$ with density for T=50 MeV 
is dominated by the increase of $s$, leading to the value of
$\eta/s$ being smaller than for the higher temperatures.
However, as the density is further increased, $\eta/s$ remains
almost constant for T=50 MeV, whereas there is a slow decrease
for the higher temperatures.
In chiral SU(3) model, the values of $\eta/s$ are obtained as
0.69 (0.73), 0.66 (0.69) and 0.66 (0.67) for 
T=50, 100 and 150 MeV for density $\rho_B=\rho_0$ 
and 
0.49 (0.51), 0.25 (0.27) and 0.22 (0.22) 
for density $\rho_B=4\rho_0$, 
for symmetric (asymmetric with $\eta_N$=0.6) 
nuclear matter. 
In symmetric nuclear matter, for T=50 MeV
and $\rho_B=\rho_0 (4\rho_0)$, the values calculated are
0.69 (0.32) for the free nucleon gas.
For the higher values of temperature T=100 and 150 MeV, 
the values of $\eta/s$ are observed to be 
smaller in the chiral SU(3) model as compared to the values for
free nucleon gas. However, for the smaller temperature of T=50 MeV,
the medium modifications of the nucleons using chiral SU(3) model
is observed to lead to higher values of $\eta/s$ at densities 
larger than $\rho_B=\rho_0$, as compared to the free nucleon gas 
due to the smaller value of the entropy density
(see figure \ref{Dens_Entr}), which dominates over 
the modification of the shear viscosity, $\eta$,
giving rise to a larger value for the ratio $\eta/s$
in the chiral SU(3) model as compared to the free nucleon gas.  
The isospin asymmetry is observed to lead to larger values
of $\eta/s$ as compared to isospin symmetric matter,
both for free nucleon gas and the chiral SU(3) model.
However, its effect is observed to be very small for the higher
values of temperature T=100 and 150 MeV as compared to
the lower value of temperature T=50 MeV. 

The effects of the baryon density and the isospin asymmetry 
of the medium on the thermal conductivity, $\kappa$ 
(in units of ${\rm fm}^{-2}$) are shown in 
figure \ref{Dens_therm_cond} for different values of the
temperature for free nucleon gas and the chiral SU(3) model. 
For free nucleon gas (shown in (a)), 
for T=50 and 100 MeV, the values of thermal conductivity 
are observed to increase with density, but the variation 
is extremely slow, with values of around 0.17 (0.2)  and
0.23 (0.24) at baryon density 
$\rho_0 (4\rho_0)$,
for T=50 and 100 MeV respectively. For T=150 MeV, there
is observed to be an initial drop in $\kappa$ with increase in density,
up to around $\rho_0$, beyond which the value is observed to be
almost constant.
In the chiral SU(3) model (shown in subplot (b)),
$\kappa$ is observed to increase with density
for T=50 and 100 MeV, both for symmetric and asymmetric 
nuclear matter, with the value observed to be higher 
in the presence of isospin asymmetry.
On the other hand, the value of $\kappa$ is observed
to decrease sharply with density up to around $\rho_B=\rho_0$,
followed by a slow drop at higher densities.
The observed density dependence of the thermal conductivity  
for free nucleon gas and the chiral SU(3) model can be understood
from the contributions from the relaxation time $\tau_i$ 
and the expression of $\kappa$ given by equation (\ref{kappa})
(modulo $\tau_i$) as follows. For the free nucleon gas,
in the absence of the nucleon medium effects, 
the drop in the $\tau_i$ with density is similar to the rise
in the latter, which is observed as the value of $\kappa$
to be almost constant at high densities.
However, the multiplying factor 
$c_f\equiv \frac{1}{3} \Big(\frac {\cal W}{\rho_B T}\Big)^2$
has a larger contribution at small densities due to
the inverse quadratic power dependence on $\rho_B$.
In the chiral SU(3) model, for T=50 and 100 MeV,
the integral (modulo $\tau_i$) multiplied by
the factor $c_f$ has an increase with density,
which dominates over the drop in the relaxation time,
which is observed as a rise of $\kappa$ with density. 
The effect is observed to be much larger for T=50 MeV
as compared to T=100 MeV. At T=150 MeV, there is observed
to be a drop with density due to the decrease in
the relaxation time, which dominates over the increasing
contribution of the integral (modulo $\tau_i$) multiplied by $c_f$.
The effects due to isospin 
asymmetry are observed to lead to higher values
of the thermal conductivity, and, the effect is 
is observed to be dominant in the chiral SU(3) model,
as compared to the free nucleon gas.
In the chiral SU(3) model, 
the values of the coefficient of thermal conductivity, 
$\kappa$ (in units of ${\rm fm}^{-2}$) are obtained as
0.21 (0.25), 0.27 (0.29) and 0.71 (1.46) for $\rho_B=\rho_0$
and 0.61 (1.12), 0.44 (0.57) and 0.67 (1.2) 
at the higher density of $\rho_B=4\rho_0$
for T=50, 100 and 150 MeV respectively 
for symmetric (asymmetric with $\eta_N$=0.6) 
nuclear matter. 
At the lower value of the
temperature T=50 MeV, the increase in the value of $\kappa$ 
is observed to be significant at high densities, with 
appreciable contributions from isospin asymmetry of the medium.
For symmetric nuclear matter, at baryon density  
$\rho_B=\rho_0 (4\rho_0)$ and T=100 MeV, 
the value of the thermal conductivity $\kappa$ (in fm$^{-2}$), 
obtained to be 0.27 (0.44) in the chiral SU(3) model, 
may be compared to the value of 
0.22 (0.23) 
of Ref. \cite{Danielewicz_PLB_146_168_1984}
and of around 0.18 (0.2) in Ref. 
\cite{NPA573_554_1994_Monras_Trans_coeffs_Nucl_Neutron_matter}.
The results of Ref. 
\cite{NPA573_554_1994_Monras_Trans_coeffs_Nucl_Neutron_matter}
were observed
to be very similar to the results using the relaxation
time approximation in Ref. \cite{Hakim_Mornas_PRC47_2846_1993}.
The values of the thermal conductivity calculated 
within the chiral SU(3) model at high densities 
are observed to be larger as compared to the values
obtained in earlier Refs. 
\cite{Danielewicz_PLB_146_168_1984,NPA573_554_1994_Monras_Trans_coeffs_Nucl_Neutron_matter,Hakim_Mornas_PRC47_2846_1993}.
The isospin asymmetry leads to higher values of $\kappa$
as compared to isospin symmetric nuclear matter,
and, the isospin asymmetry effects are observed to be
quite significant. These can have consequences, e.g., 
for the collective flow of the hadrons in the CBM experiment
at FAIR at the future facility of GSI, where, highly 
compressed baryonic matter are planned to be produced
\cite{TheCBMPhysicsBook}.
\section{Summary}
To summarize, we have studied the thermodynamic
properties and transport coefficients
of the shear viscosity and thermal conductivity
in hot asymmetric nuclear matter using a chiral SU(3) model. 
The masses and chemical potentials of the
nucleons are modified in the medium due to their interactions
with the mean scalar and vector fields within the model. 
The transport coefficients are obtained from solution
of the Boltzmann equation using the first order departure 
from the local equilibrium distribution function. 
These are calculated within the relaxation time approximation.
The transport coefficients depend on the effective masses of the nucleons 
and the mean fields in the hot asymmetric nuclear medium, 
calculated within the chiral SU(3) model. The temperature
and density dependent relaxation times of the nucleons 
are calculated from their mean free paths divided by 
the mean velocities of these particles in the thermal medium.
There is observed
to be a difference in the relaxation times for the proton and
neutron in the asymmetric matter arising from their 
equilibrium distribution functions, which are different
due to the difference in their
in-medium masses and effective chemical potentials.
For a given temperature, the shear viscosity coefficient 
to entropy density ratio, $\eta/s$ is observed to drop with
increase in density. The shear viscosity, $\eta$,
calculated within the chiral SU(3) model is observed 
to be smaller than the values calculated for the free nucleon gas,
whereas, the thermal conductivity, $\kappa$ is observed to
be appreciably larger as compared to the free nucleon gas.
The effect of isospin asymmetry is observed to lead 
to higher values of the coefficient of shear viscosity 
to entropy density ratio ($\eta/s$)
and the coefficient of thermal conductivity. 
The modifications of $\eta$ due to isospin effects are observed 
to be marginal, whereas, the effect of the isospin 
asymmetry of the medium on the coefficient 
of thermal conductivity is observed to lead to 
appreciably larger values as compared to isospin
symmetric nuclear matter for higher values 
of the temperature.
For T=150 MeV, the coefficient of thermal conductivity is observed 
to drop with increase in density, contrary to the 
rise observed for the lower values of temperature, 
T=50 and 100 MeV.
The present study can be of relevance to the experimental
observables, e.g., the collective flow, and hadron 
spectra in CBM experiment at FAIR at the future facility
of GSI. 

\acknowledgements
A.M. would like to acknowledge the kind hospitality
at Institut f\"ur Theoretische Physik, University
of Frankfurt, where the work was initiated.
A.M. also acknowledges financial support from I.I.T. Delhi
for the visit.


\begin{thebibliography}{100}

\bibitem{Danielewicz_PLB_146_168_1984}
P. Danielewicz, Phys. Lett. B {\bf 146}, 168 (1984).

\bibitem{Prakash_Phys_Rep_227_321_1993}
M. Prakash, M. Prakash, R. Venugopalan and G. Welke, 
Phys. Rep. {\bf 227}, 321 (1993).

\bibitem{Itakura_PRD77_014014_2008} 
K. Itakura, O. Morimatsu and H. Otomo, Phys. Rev. D {\bf 77},
014014 (2008).

\bibitem{Recent_Prog_QHD_Serot_Walecka_IJMPE6_515_1996}
B.D. Serot and J. D. Walecka, Int. Jour. Mod. Phys. E {\bf 6},
515 (1997).

\bibitem{NPA573_554_1994_Monras_Trans_coeffs_Nucl_Neutron_matter}
L. Mornas, Nucl. Phys. A {\bf 573}, 554 (1994).
\bibitem{Hakim_Mornas_PRC47_2846_1993}
R. Hakim and L. Mornas, Phys. Rev. C {\bf 47}, 2846 (1993).

\bibitem{Ayik_Ivanov_Russkikh_Norenberg_NPA578_640_1994}
S. Ayik, Y. B. Ivanov, V. N. Russkikh and W. N\"orenberg,
Nucl. Phys. A {\bf 578}, 640 (1994).

\bibitem{Abu_Samreh_NPA552_1993_101}
M. M. Abu-Samreh and H. S. K\"ohler, Nucl. Phys. A
{\bf 552}, 101 (1993).

\bibitem{A_S_Khvorostukhin_NPA915_198_2013}
A. S. Khvorostukhin, V. D. Toneev and D. N. Voskresensky,
Nucl. Phys. A {\bf 915}, 198 (2013).

\bibitem{Chakraborty_Kapusta_PRC83_014906_2011}
P. Chakraborty and J. I. Kapusta, Phys. Rev. C {\bf 83},
014906 (2011).

\bibitem{Albright_Kapusta_PRC93_014903_2016}
M. Albright and J. I. Kapusta, Phys. Rev. C {\bf 93},
014903 (2016).

\bibitem{PRC86_024913_2012}
J. Noronha-Hostler, J. Noronha and C. Greiner, Phys. Rev. C {\bf 86},
024913 (2012).

\bibitem{PRC77_024911_2008_Gorenstein}
M. I. Gorenstein, M. Hauer and O. N. Moroz, Phys. Rev. C
{\bf 77}, 024911 (2008).

\bibitem{Phys_Rev_D31_53_1985_Danielewicz_Gyulassy}
P. Danielewicz and M. Gyulassy,
Phys. Rev. D {\bf 31}, 53 (1985).


\bibitem{Hosoya_Kajantie_NPB250_666_1985}
A. Hosoya and K. Kajantie, Nucl. Phys. B {\bf 250}, 666 (1985).

\bibitem{PRC84_035202_2011_visc_gluon_matter}
A. S. Khvorostukhin, V. D. Toneev and D. N. Voskresensky,
Phys. Rev. C {\bf 84}, 035202 (2011).

\bibitem{Gavin_NPA435_1985_826}
S. Gavin, Nucl. Phys. A {\bf 435}, 826 (1985).

\bibitem{A_S_Khvorostukhin_NPA845_106_2010}
A. S. Khvorostukhin, V. D. Toneev and D. N. Voskresensky,
Nucl. Phys. A {\bf 845}, 106 (2010).

\bibitem{Phys_Rev_C103_054901_2021_Elena}
O. Soloveva, D. Fuseau, J. Aichelin, E. Bratkovskaya,
Phys. Rev. C {\bf 103}, 054901 (2021).
\bibitem{2408_00524_Isabella_Moore}
I. Danhoni and G. D. Moore, J. High Energ. Phys. 2024, 75 
(2024).
\bibitem{Hybrid_HIC}
A. Sch\"afer, I. Karpenko, X.-Y. Wu, J. Hammelmann and
H. Elfner, Eur. Phys. Jour. A {\bf 58}, 230 (2022).
\bibitem{Hybrid_HIC_shear_1}
I. A. Karpenko, P. Huovinen, H. Peterson and M. Bleicher,
Phy. Rev. C {\bf 91}, 064901 (2015). 
\bibitem{Gotz_Hanah_PRC106_054904_2022}
N. G\"otz and H. Elfner, Phys. Rev. C {\bf 106}, 054904 (2022).
\bibitem{Gotz_Hanah_2503_10181}
N. G\"otz, I. Karpenko and H. Elfner, 
Phys. Rev. C {\bf 112}, 014910 (2025).

\bibitem{KSS}
P. Kovtun, D. T. Son and A. O. Starinets, Phys. Rev. Lett.
{\bf 94}, 111601 (2005).

\bibitem{particle_flow_anisotropies_1}
B. Schenke, S. Jeon and C. Gale, Phys. Rev. Lett. {\bf 106},
042301 (2011): ibid. Phys. Rev. C {\bf 85}, 024901 (2012).
\bibitem{particle_flow_anisotropies_2}
H. Song, S. A. Bass and U. Heinz, Phys. Rev. C {\bf 83},
054912 (2011)
\bibitem{particle_flow_anisotropies_3}
H. Song, S. A. Bass, U. Heinz, T. Hirano and C. Shen,
Phys. Rev. C {\bf 83}, 054910 (2011); ibid, Phys. Rev. Lett. 
{\bf 106}, 192301 (2011).

\bibitem{BUU_1}
 B.Bl\"attel, V. Koch, W. Cassing, U. Mosel, 
Phys. Rev. C {\bf 38}, 1767 (1988).
\bibitem{BUU_2}
A. Lang, B.Bl\"attel, V. Koch, W. Cassing, U. Mosel, 
PLB245, 147 (1990).
\bibitem{BUU_3}
 T. Maruyama, B.Bl\"attel, W. Cassing, A. Lang, U. Mosel, 
K. Weber, Phys. Lett. B {\bf 297}, 228 (1992).
\bibitem{BUU_4}
T. Maruyama, W. Cassing, U. Mosel, S. Teis,
K. Weber, Nucl. Phys. A {\bf 573}, 653 (1994).
\bibitem{BUU_5}
B.Bl\"attel, V. Koch, K. Weber, W. Cassing, U. Mosel, 
Nucl. Phys. A {\bf 495}, 381c (1989).
\bibitem{BUU_6}
B.Bl\"attel, V. Koch, A. Lang, W. Cassing, U. Mosel, 
Phys. Rev. C {\bf 43}, 2728 (1991).
\bibitem{BUU_7}
V. Koch, B.Bl\"attel, W. Cassing, U. Mosel, 
Phys. Lett. B {\bf 241}, 174 (1990).
\bibitem{BUU_8}
 V. Koch, B.Bl\"attel, W. Cassing, U. Mosel, 
Nucl. Phys. B {\bf 532}, 715 (1991).

\bibitem{paper3}
 	P. Papazoglou, D. Zschiesche, S. Schramm, J. Schaffner-Bielich,
	H. St\"ocker, and W. Greiner, Phys. Rev. C {\bf 59},  411  (1999).
\bibitem{hartree}
	D. Zschiesche, A. Mishra, S. Schramm, H. St\"ocker and W. Greiner,
        Phys. Rev. C {\bf 70}, 045202 (2004).
\bibitem{kristof1}
	A. Mishra, K. Balazs, D. Zschiesche, S. Schramm,
	H. St\"ocker, and W. Greiner,
        Phys. Rev. C {\bf 69}, 024903 (2004).
\bibitem{Schramm_2013} 
S. Schramm, V. Dexheimer, R. Negreiros, J. Steinheimer and
T. Sch\"urhoff, arXiv: 1310.5804 [astro-ph.SR].
\bibitem{Dex_2015} 
V. Dexheimer, R. Negreiros and S. Schramm, Phys. Rev. C {\bf 91},
055808 (2015).
\bibitem{JSB_2016} 
A. Zacchi, M. Hanauske and J. Schaffner-Bielich, Phys. Rev. D
{\bf 93}, 025001 (2016).
\bibitem{AMSPMWG_2015}
Amruta Mishra, S. P. Misra and W. Greiner, Int. Jour. Mod. Phys.
E {\bf 24}, 1550053 (2015).
\bibitem{AMSPM_2017}
A. Mishra and S. P. Misra, Phys. Rev. C {\bf 95}, 065206 (2017).
\bibitem{AMSPM_DW_HQ_DS_PV_2023} 
Amruta Mishra and S. P. Misra, Phys. Rev. D {107}, 074003 (2023). 
\bibitem{AMAKSPM24} 
Amruta Mishra, Arvind Kumar and S. P. Misra, Phys. Rev. D {110}, 
014003 (2024).

\bibitem{TheCBMPhysicsBook}
The CBM Physics Book, Lect. Notes Phys. {\bf 814}, 
Springer-Verlag Berlin Heidelberg 2010, B. Friman et al (Eds.)
\bibitem{P_Senger_HIC_FAIR_NICA_Energies}
P. Senger,
Particles 2021,
4, 214-226, https://doi.org/10.3390/particles4020020.
\bibitem{BES_RHIC}
A. Bzdak, S. Esumi, V. Koch, J. Liao, M. Stephanov,
and N. Xu, 
Phys. Rept. 853, 1 (2020).
\bibitem{Alford_NS}
Mark G. Alford, L. Bovard, M. Hanauske, L. Rezzolla and
K. Schwenzer, Phys. Rev. Lett. {\bf 120}, 041101 (2018).

\bibitem{Weinberg}
S.Weinberg, Phys. Rev. {\bf 166} 1568 (1968).
\bibitem{coleman}
S. Coleman, J. Wess, B. Zumino, Phys. Rev. {\bf 177} 2239 (1969);
C.G. Callan, S. Coleman, J. Wess, B. Zumino, Phys. Rev. {\bf 177}
2247 (1969).
\bibitem{Bardeen}
W. A. Bardeen and B. W. Lee, Phys. Rev. {\bf 177} 2389 (1969).

\bibitem{sche1} J. Schechter, Phys. Rev. D {\bf 21}, 3393 (1980).
 \bibitem{heide1}
Erik K. Heide, Serge Rudaz and Paul J. Ellis, Nucl. Phys. A {\bf 571}, 
(1994) 713.

\bibitem{AM_AK_PRC81_065204_2010}
Arvind Kumar and Amruta Mishra, Phys. Rev. C {\bf 81}, 065204 (2010).
\bibitem{Li_Ko}
G. Q. Li, C. M. Ko and G. E. Brown, Nucl. Phys. A {\bf 606}, 568
(1996).
\bibitem{Furnstahl_Serot}
R. J. Furnstahl and B. D. Serot, Phys. Rev. C {\bf 41},
262 (1990).

\end{thebibliography}
\end{document}